\documentclass[11pt]{article}
\usepackage{amsfonts,amsmath,amssymb,bm,bbm,enumerate,graphicx,color}
\usepackage{subfigure}
\usepackage{epstopdf}
\usepackage[dvipsnames]{xcolor}
\usepackage{graphicx}			
\usepackage{mathrsfs}
\usepackage{amsfonts}
\usepackage{algorithm}
\usepackage{algorithmic}
\usepackage[dvipsnames]{xcolor}
\usepackage[numbers]{natbib}

\usepackage[normalem]{ulem}
\usepackage[colorlinks=true,urlcolor=purple,citecolor=blue,linkcolor=red,bookmarks=true]{hyperref}
\usepackage{tikz}
\usepackage{multicol}
\usepackage{multirow}
\usepackage{enumitem}
\usepackage[T1]{fontenc}
\usepackage[utf8]{inputenc}
\usepackage{array}
\usepackage{booktabs}

\numberwithin{The}{section}
\allowdisplaybreaks

\newcounter{quest}

\usepackage{footnote}
\usepackage[affil-it]{authblk}
\usepackage[english]{babel}

\def\boxit#1{\vskip 1ex \hrule\hbox{\vrule\kern6pt\vbox{\vskip 6pt#1\vskip 6pt}\kern6pt\vrule}\hrule}

\begin{document}
\title{\bf Bayesian Nowcasting Data Breach IBNR Incidents}
\author[a]{Maochao Xu \thanks{Corresponding author. Email: mxu2@ilstu.edu.}}
\author[b]{Hong Sun}
\author[b]{Peng Zhao}
\newcommand\CoAuthorMark{\footnotemark[\arabic{footnote}]}
\affil[a]{\rm Department of Mathematics, Illinois State University, USA}
\affil[b]{\rm School of Mathematics and Statistics, Jiangsu Normal University, China}
\maketitle
\titlepage
\begin{abstract}
The reporting delay in data breach incidents poses a formidable challenge for Incurred But Not Reported (IBNR) studies, complicating reserve estimation for actuarial professionals. This work presents a novel Bayesian nowcasting model designed to accurately model and predict the number of IBNR data breach incidents. Leveraging a Bayesian modeling framework, the model integrates time and heterogeneous effects to enhance predictive accuracy. Synthetic and empirical studies demonstrate the superior performance of the proposed model, highlighting its efficacy in addressing the complexities of IBNR estimation. Furthermore, we examine reserve estimation for IBNR incidents using the proposed model, shedding light on its implications for actuarial practice.	
	
\end{abstract}

\section{Introduction}
Data breaches pose a significant threat to computer systems and have become an ongoing problem due to the extensive network activities. Numerous severe cybersecurity incidents have demonstrated the gravity of this issue. For instance, the Privacy Rights Clearinghouse (PRC) has reported over 11 billion records breached since 2005. In 2022, the Identity Theft Resource Center (ITRC) recorded 1,802 data breach incidents, affecting approximately 422,143,312 individuals. Similarly, in 2021, there were 1,862 incidents reported by the ITRC, impacting around 293,927,708 individuals\footnote{\url{https://www.idtheftcenter.org/publication/2022-data-breach-report/}}.
The cost associated with data breaches is also considerable. According to the Cost of a Data Breach Report 2023 published by IBM\footnote{\url{https://www.ibm.com/reports/data-breach}} \footnote{The Cost of a Data Breach Report 2023, conducted independently by Ponemon Institute and sponsored, analyzed, and published by IBM Security, examined 553 organizations affected by data breaches occurring between March 2022 and March 2023.}, the average cost of a data breach incident increased from \$4.35 million in 2022 to \$4.45 million in 2023. Additionally, the average cost per breached record saw a 0.6\% increase from 2022 to 2023, reaching \$165 compared to \$164. Taking a long-term perspective, the average total cost of a data breach has risen by 15.3\% since 2020, when it was \$3.86 million.

The unique nature of cyber risk often leads to the discovery of data breaches days, months, or even years after the initial breach occurred. In the 2023 IBM report, it is stated that the mean time to identify a data breach from 2021 to 2023 was 212, 207, and 204 days, respectively. Similarly, the mean time to contain a data breach was 287, 277, and 277 days, respectively. As breaches go unaddressed for longer periods, more data is leaked, resulting in a larger overall impact, both financially and otherwise. For instance, in 2023, a breach with a lifecycle exceeding 200 days had an average cost of \$4.95 million, whereas breaches with a lifecycle of less than 200 days cost an average of \$3.93 million. This poses a significant challenge for estimating reserves for incurred but not reported (IBNR) breach incidents. Solvency II regulations require reserves to be set based on the best estimate of the loss distribution, which further complicates the IBNR study.  It should be noted that there are many studies on the IBNR in the actuarial and insurance literature \cite{bornhuetter1972actuary,wuthrich2008stochastic,wuthrich2018machine,delong2022collective}. However, there is limited research on IBNR studies for cyber data breach incidents in the existing literature. One of the most relevant works in this area is by Xu and Nguyen  \cite{xu2022statistical}, where they developed a novel statistical approach to model the time to identify an incident and the time to notice affected individuals. However, their study did not specifically address the IBNR issue, which is the focus of the current study.

Other studies offer tangential insights into data breach dynamics.   For example, Buckman et al.  \cite{buckman2017organizations} studied the time intervals between data breaches for enterprises that had at least two incidents between 2010 and 2016. They showed that the duration between two data breaches might increase or decrease, depending on some factors. Edwards et al. \cite{edwards2016hype} analyzed the temporal trend of data breach size and frequency and showed that the breach size follows a log-normal distribution and the frequency follows a negative binomial distribution. They further showed that the frequency of large breaches (over 500,000 breached records) follows the Poisson distribution, rather than the negative binomial distribution, and that the size of large breaches still follows a log-normal distribution. Eling and Loperfido  \cite{eling2017data} studied data breaches from actuarial modeling and pricing perspectives. They used multidimensional scaling and goodness-of-fit tests to analyze the distribution of data breaches. They showed that different types of data breaches should be analyzed separately and that breach sizes can be modeled by the skew-normal distribution. Sun et al. \cite{sun2021modeling} developed a frequency-severity actuarial model of aggregated enterprise-level breach data to promote ratemaking and underwriting in insurance, and they also developed a 
multivariate frequency-severity model to examine the status of data breaches at the state level in \cite{sun2023multivariate}. Ikegami and Kikuchi  \cite{ikegami2020modeling} studied a breach dataset in Japan and developed a probabilistic model for estimating the data breach risk. They showed that the inter-arrival times of data breaches (for those enterprises with multiple breaches) follow a negative binomial distribution. Romanosky et al.  \cite{romanosky2011data} used a fixed effect model to estimate the impact of data breach disclosure policy on the frequency of identity thefts incurred by data breaches. The proposed study is significantly different from those in the literature since we aim to address the important yet unstudied problem–IBNR cyber incidents. %{\color{blue}Further, we discuss the required reserve for IBNR incidents in a policy period.}

Nowcasting, or predicting the present, is an effective method for estimating the  IBNR  events in real-time, using an incomplete set of reports. This approach has been widely utilized in various fields, including disease surveillance and weather forecasting \cite{mcgough2019nowcasting,bastos2019modelling}. By leveraging available data and statistical techniques, nowcasting enables analysts to estimate the current state of events that may not have been fully reported or observed yet. This methodology can be particularly useful in estimating the number of IBNR events in the context of data breaches, where the reporting of incidents may be delayed or incomplete. In this study, we develop a novel Bayesian nowcasting approach to modeling and predicting the number of IBNR data breach incidents within one year. By examining temporal and heterogeneous effects on IBNR modeling, this research aims to enhance the accuracy of reserve estimations for cyber incidents.

The rest of the paper is organized as follows. Section \ref{sec:eda} introduces the data, and performs the exploratory data analysis. In Section
\ref{sec:model}, we introduce the proposed Bayesian nowcast model. Section \ref{sec:syn} examines the model performance on synthetic data. In Section \ref{sec:empirical}, we employ the proposed model to the empirical data and present the reserve estimation. The last Section concludes the current work and present some discussion.  

\section{Data Preprocessing  and Exploratory Data Analysis}\label{sec:eda}
The incident data was collected from the ITRC, which gathers information about publicly reported data breaches from various sources, including company announcements, mainstream news media, government agencies, recognized security research firms and researchers, and non-profit organizations. We focus on recent incidents with breach dates from 1/1/2018 to 12/31/2023, and drop the incidents without the information of affected individuals. After further removing the duplicates, we have a total of 859 incidents.

\begin{figure}[htbp!]
	\centering
	\includegraphics[width=0.7\textwidth]{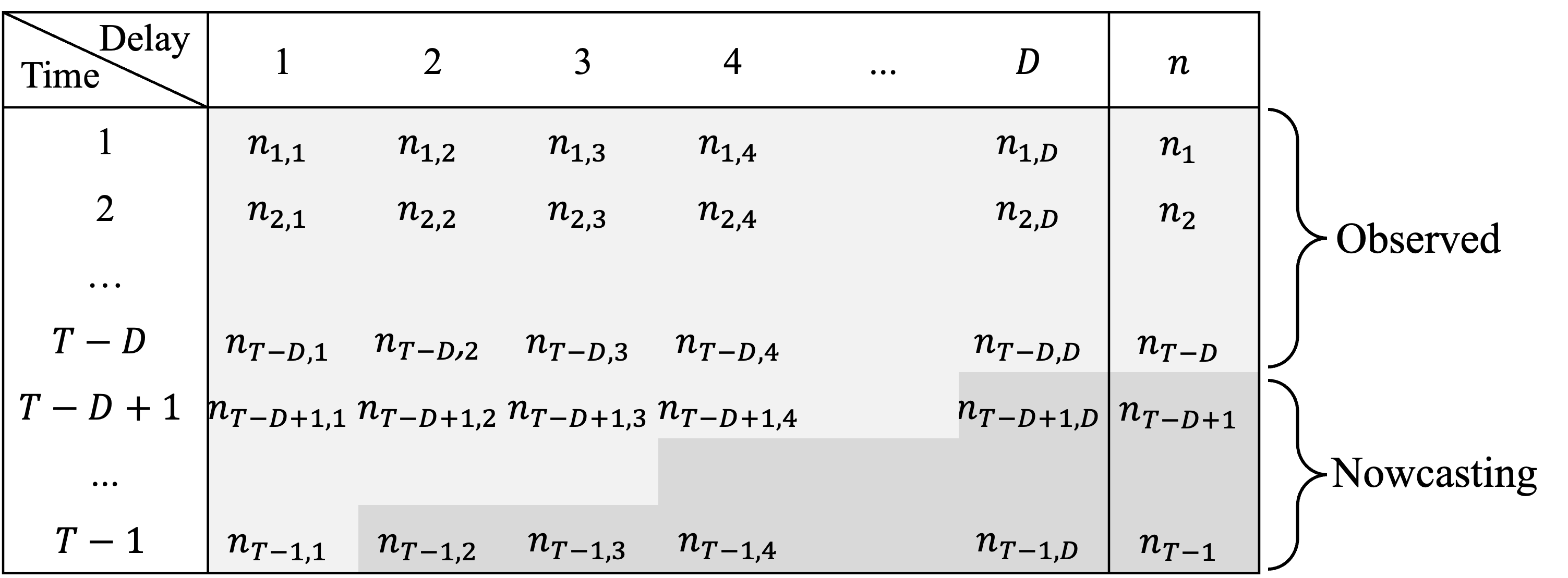}
	\caption{Data structure in the reporting delay of cyber incidents. The values in the gray area represent observed data, and the values in the gray area represent the number of incurred but not reported (IBNR) events. Unit: month.\label{fig:delay-triangle}}
\end{figure}

Figure \ref{fig:delay-triangle} shows the data structure for the reporting delay of cyber incidents based on the breach data. The rows correspond to time $t = 1, 2,\ldots,T$, and $T$ is the present time. The columns correspond to the amount of delay in the same unit as $t$. In our data, the unit is set to be one month. $n_{t,D}$ represents the number of incidents was reported between $(D-1)$th and $D$th months after occurrences at time $t$. For example, for the first row, $n_{1,1}$ represents the number of breach incidents that occurred during the first month of our study period and were reported within the month after they occurred,  $n_{1,2}$ incidents were reported between the second month after occurrence, and $n_{1,3}$ incidents were reported between the second and third month after occurrence. In our study, we chose $D=12$ because one year is a typical choice for the reserving period \cite{wuthrich2008stochastic}. That is, the incidents with a delay larger than one year were excluded from the current study. The last column $n_t$ represents the total incidents occurring at time period $t$. Since the current time is $T$, the completely observable times are from 1 to $T-D$. From $T-D+1$ to $T-1$, we can only observe partial observations and need to nowcast the incidents (i.e., determining the number of IBNR incidents). %The values from $T$ to $T+K$ are future events that we are interested in forecasting.

 The dataset studied is divided into 72-month time periods. For the purpose of assessing model performance, we use the first 60 months as the in-sample data (i.e., 1/1/2018 to 12/31/2022), and the rest of the data as the out-of-sample data (i.e. 1/1/2023 to 12/31/2023), i.e., $T=61$.  
 \begin{figure}[htbp!]
	\centering
	\subfigure[Time series]
{\includegraphics[width=0.44\textwidth]{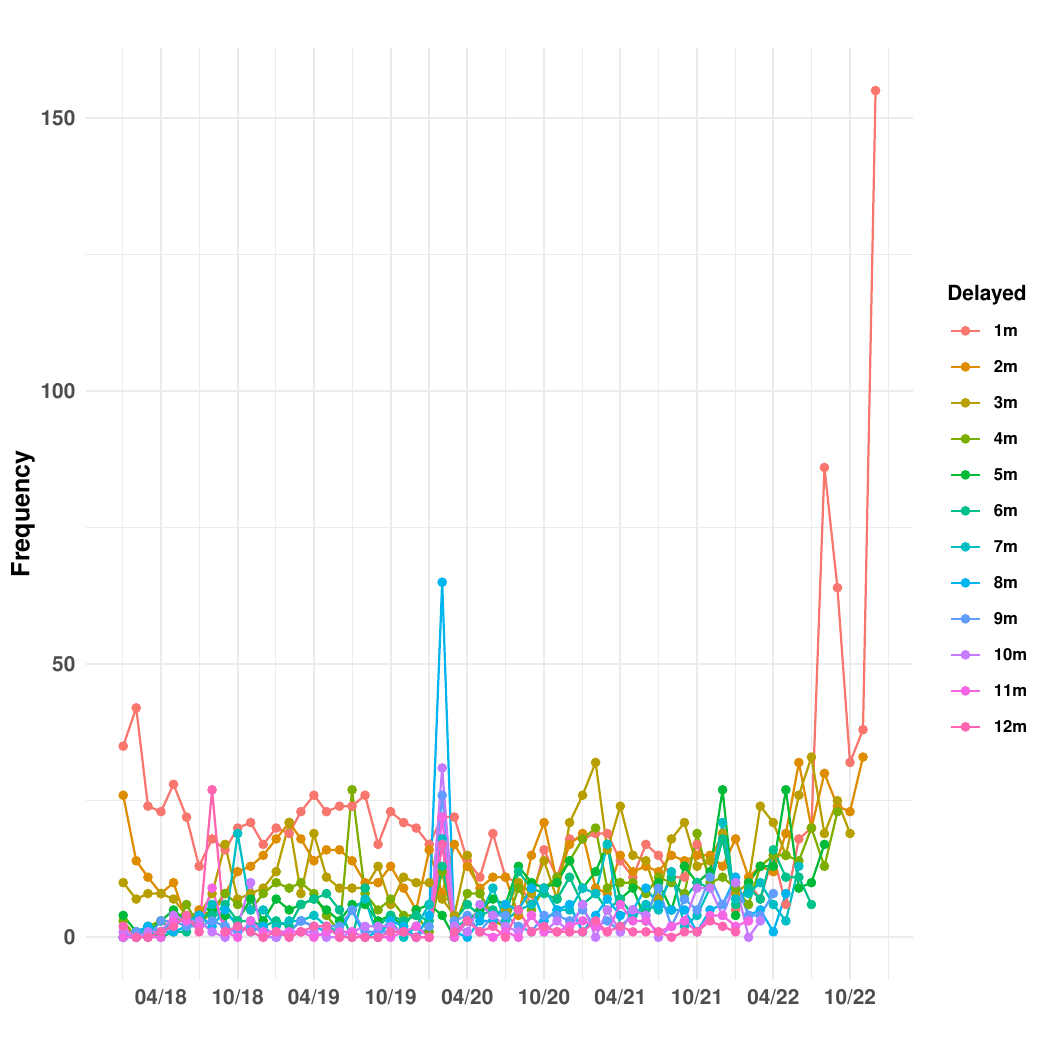}\label{fig:delay-freq1}}
	\subfigure[Boxplots]
{\includegraphics[width=0.44\textwidth]{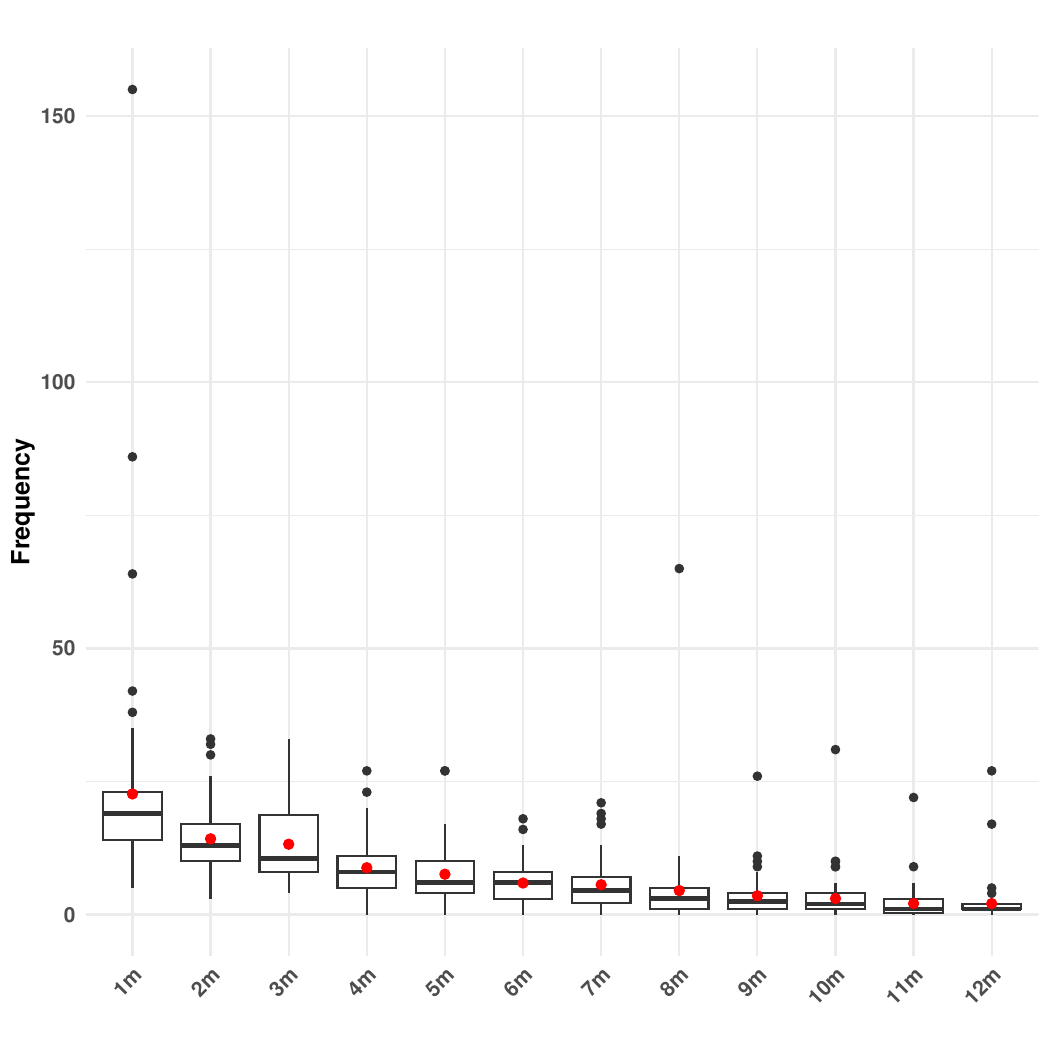}\label{fig:delay-freq2}}
	\caption{Time series plots and boxplots of frequencies of the number of data breaches reported with different delay periods. The red dot in the boxplot represents the mean.\label{fig:delay-freq}}
\end{figure}
Figure \ref{fig:delay-freq1} illustrates the time series plots of the frequency of the number of data breaches reported with different delay periods for the in-sample data. The number of incidents in different delayed months changes over time, showing an increasing trend in recent months. Particularly, we see a jump in December 2022, (i.e., 155 incidents occurred and were reported in this month). Most of the incidents were delayed in the first three months. Figure \ref{fig:delay-freq2} further shows the boxplots of frequency in various delayed months. It is clearly seen that the frequency shows a decreasing trend over delayed months. This suggests that the time effect needs to be incorporated into the modeling process.

\begin{figure}[htbp!]
	\centering
	\includegraphics[width=0.6\textwidth]{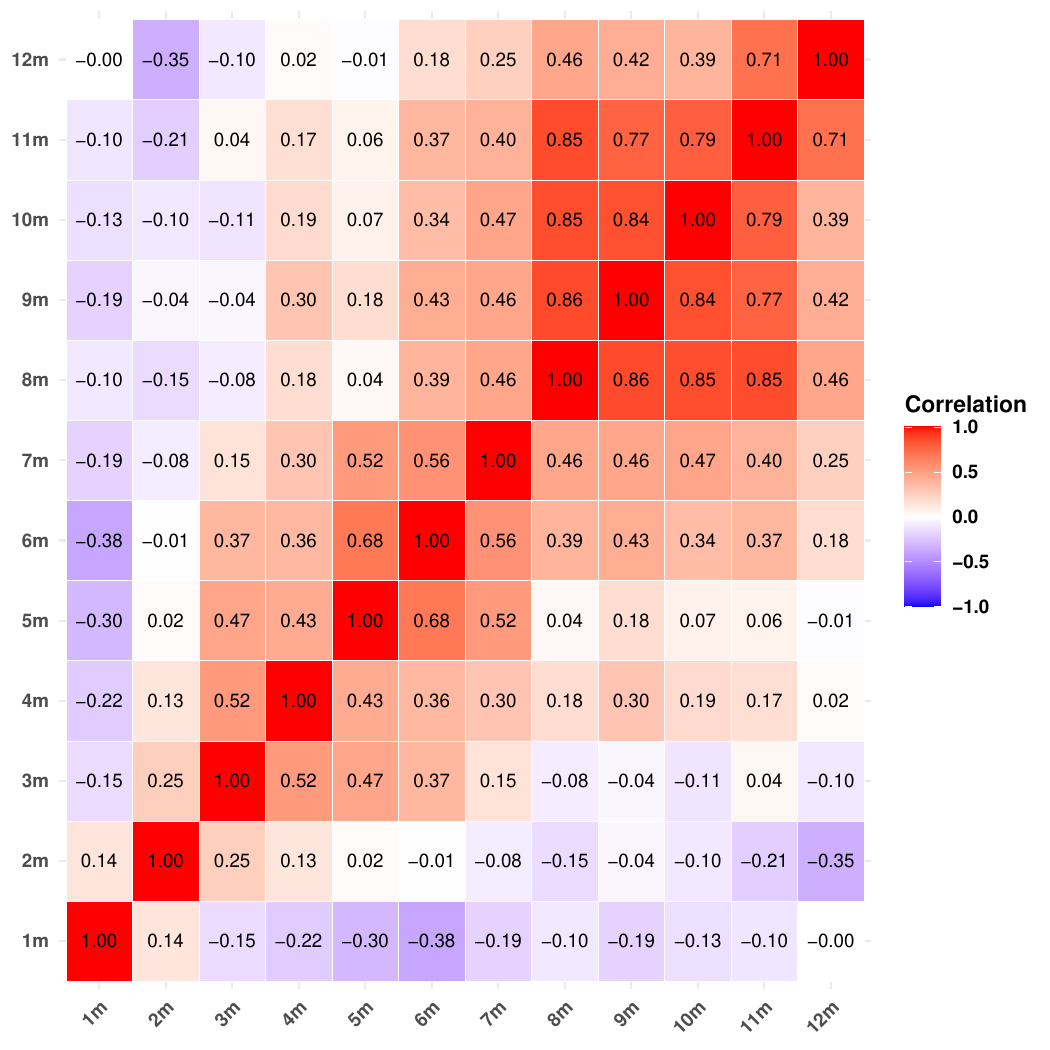}
	\caption{Correlation plots of frequencies in different delayed months.\label{fig:corplot}}
\end{figure}

 Figure \ref{fig:corplot} shows the correlation plots of the frequency among different delayed months. We observe correlations among adjacent months. It is interesting to see the high correlations among large adjacent delayed months (e.g., 10/11/12 months), which is because  
 there are a few incidents reported in those large delayed months. This suggests that the correlations among different delayed months should be considered as well.

\section{Bayesian Nowcasting Model}\label{sec:model}
The Bayesian nowcasting model proposed in this study aims to capture the time effect and correlations among the delayed months in the number of data breach events that occur but are not yet reported. Define a random variable $n_{t,d}$ that represents the number of events occurring at time $t$ (where $t = 1,2,\ldots,T-1$) but reported until time $d$ (where $d = 1,2,\ldots,D$). The model assumes that $n_{t,d}$ follows a negative binomial distribution:
\begin{equation}
	n_{t,d} \sim \text{NB}(p_{t,d},r_{t,d}),
\end{equation}
where $p_{t,d}$ is the probability of breach and $r_{t,d}$ is the number of breaches. The proposed distribution is very flexible as it allows the probability and number of branches to vary with the time and the delayed month. 
To capture the effect and correlations among the delayed months,  we propose the following Bayesian model
\begin{equation}\label{eq:logit}
{\rm logit}(p_{t,d})=\alpha_0+\alpha_1 t+\alpha_2 d/D,
\end{equation}
 and
\begin{equation}\label{eq:rtd}
	\log(r_{t,d}) = \beta_0+\beta_1 t+\beta_2 d/D,
\end{equation}
where  
\begin{equation*}
	\alpha_i \sim N\left(0,\sigma^2_{\alpha_i}\right), \quad i = 0,1,2,
\end{equation*}
and 
\begin{equation*}
	\beta_j\sim N\left(0,\sigma^2_{\beta_j} \right), \quad j = 0,1,2.
\end{equation*}
Eq. \eqref{eq:logit} models the breach probability with logistic regression with random effect effects, and Eq. \eqref{eq:rtd} models the log-transformed number of incidents with a regression model with random effects.  The prior distributions for the random effects are assumed to be normal distributions. The variability of the random effects is incorporated by assigning prior distributions, a standard exponential distribution, to the parameters $\sigma_{\alpha_i}$ and $\sigma_{\beta_j}$. That is,
$$\sigma_{\alpha_i}\sim {\rm Exp}(1),\quad \sigma_{\beta_j}\sim {\rm Exp}(1),$$
for $i,j =0,1,2$. 
The posterior distribution for the model parameters $\Theta = (\alpha_i,\beta_j, \sigma_{\alpha_i},\sigma_{\beta_j})$ can be represented as:
\begin{equation}
	p(\Theta|\mathbf{n}) = p(\Theta) \prod_{t=1}^{T-1} \prod_{d=1}^{D} p(n_{t,d}|\Theta),
\end{equation}
where $p(\Theta|\mathbf{n})$ follows the negative binomial distribution and $\mathbf{n}$ represents all the observed data. The joint prior distributions $p(\Theta)$ are specified for the model parameters. The estimation of posterior distributions is performed using  Markov Chain Monte Carlo (MCMC) methods \cite{liu2001monte,de2017programming} .

In our study, the IBNR nowcasting is done by estimating the posterior predictive distribution:
\begin{equation}
	p(n_{t,d})=\int_{\Theta} p(n_{t,d}|\Theta) p(\Theta|\mathbf{n}) d\Theta,
\end{equation}
where the MCMC approach is employed to obtain samples from the posterior predictive distribution.  

\section{Sythetic data study}\label{sec:syn}

To understand the model performance, we examine the model performance on a synthetic dataset. The data is generated from a Negative binomial distribution with 
$${\rm logit}(p_{t,d})=-1.5-.01t+.8d/D,$$
$$\log(r_{t,d})=1.5+.01t-1.8d/D,$$
where $t=1,\ldots, 72$, and $d=1,\ldots,12$, $D=12$.
\begin{figure}[htbp!]
	\centering
	\subfigure[Time series]
{\includegraphics[width=0.44\textwidth]{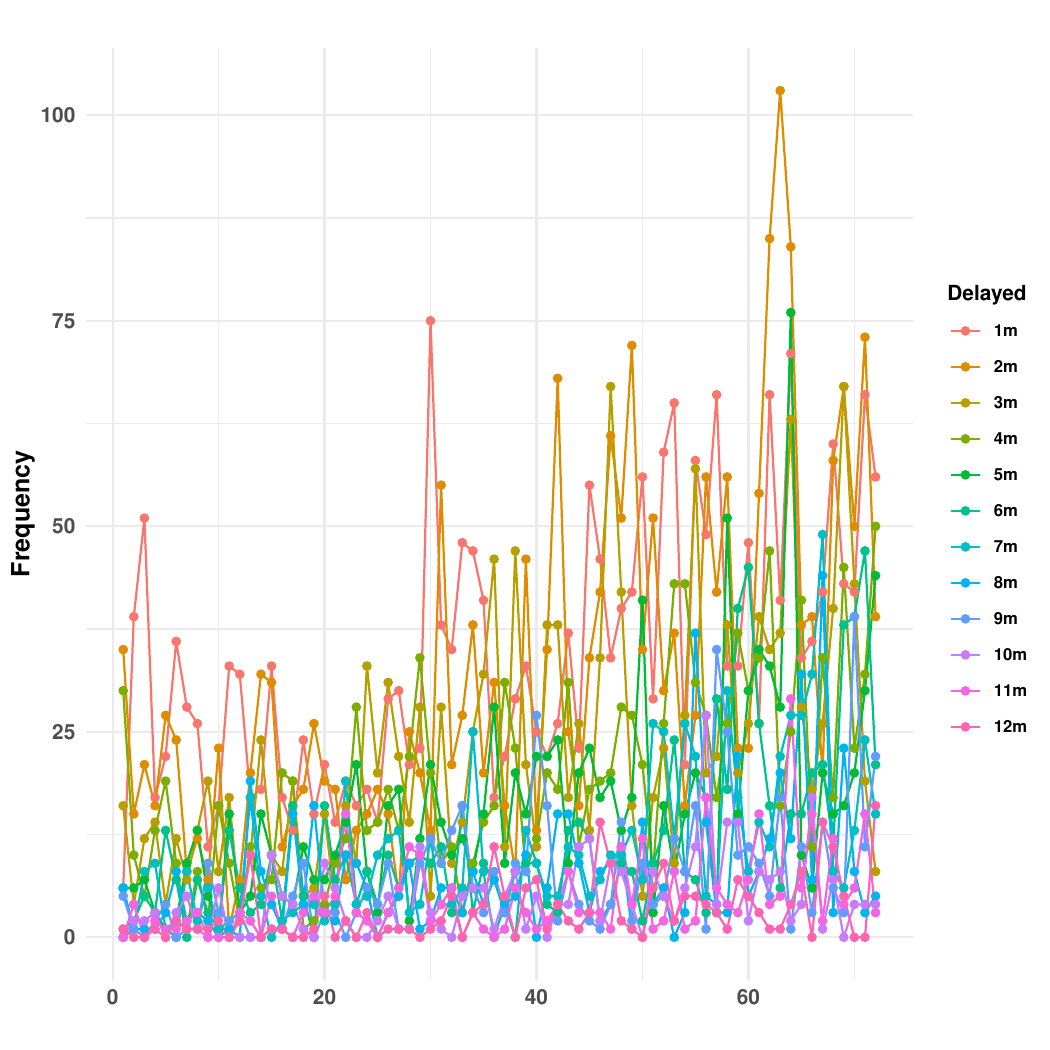}\label{fig:delay-freq1-sim}}
	\subfigure[Boxplots]
{\includegraphics[width=0.44\textwidth]{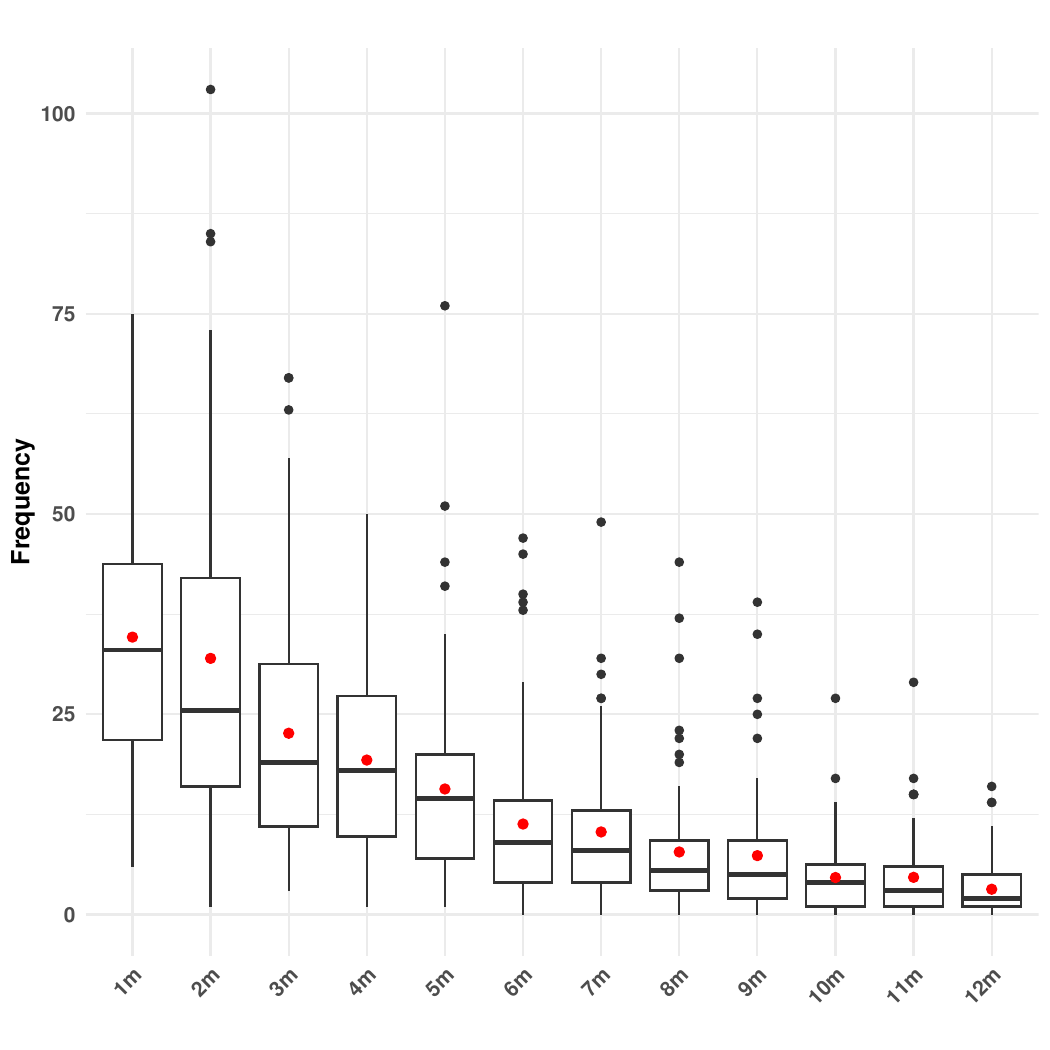}\label{fig:delay-freq2-sim}}
	\caption{Time series plots and boxplots of frequencies of the number of data breaches reported with different delay periods. The red dot in the boxplot represents the mean.\label{fig:delay-freq-sim}}
\end{figure}
The time series plots of delayed months are displayed in Figure \ref{fig:delay-freq1-sim}. It is observed that a significant portion of simulated incidents is reported within the initial months following the occurrence. The time series shows variability and an increasing trend in recent months. The boxplots in Figure   \ref{fig:delay-freq2-sim} reveal that the data is skewed, as indicated by the comparatively large means. Furthermore, there is large variability across the dataset, particularly within the initial months of delay (up to 4 delayed months). Additionally, the presence of outliers within the simulated sample further underscores the heterogeneity of incident reporting patterns.

\begin{figure}[htbp!]
	\centering
	\subfigure[$\alpha_0$]
{\includegraphics[width=0.32\textwidth]{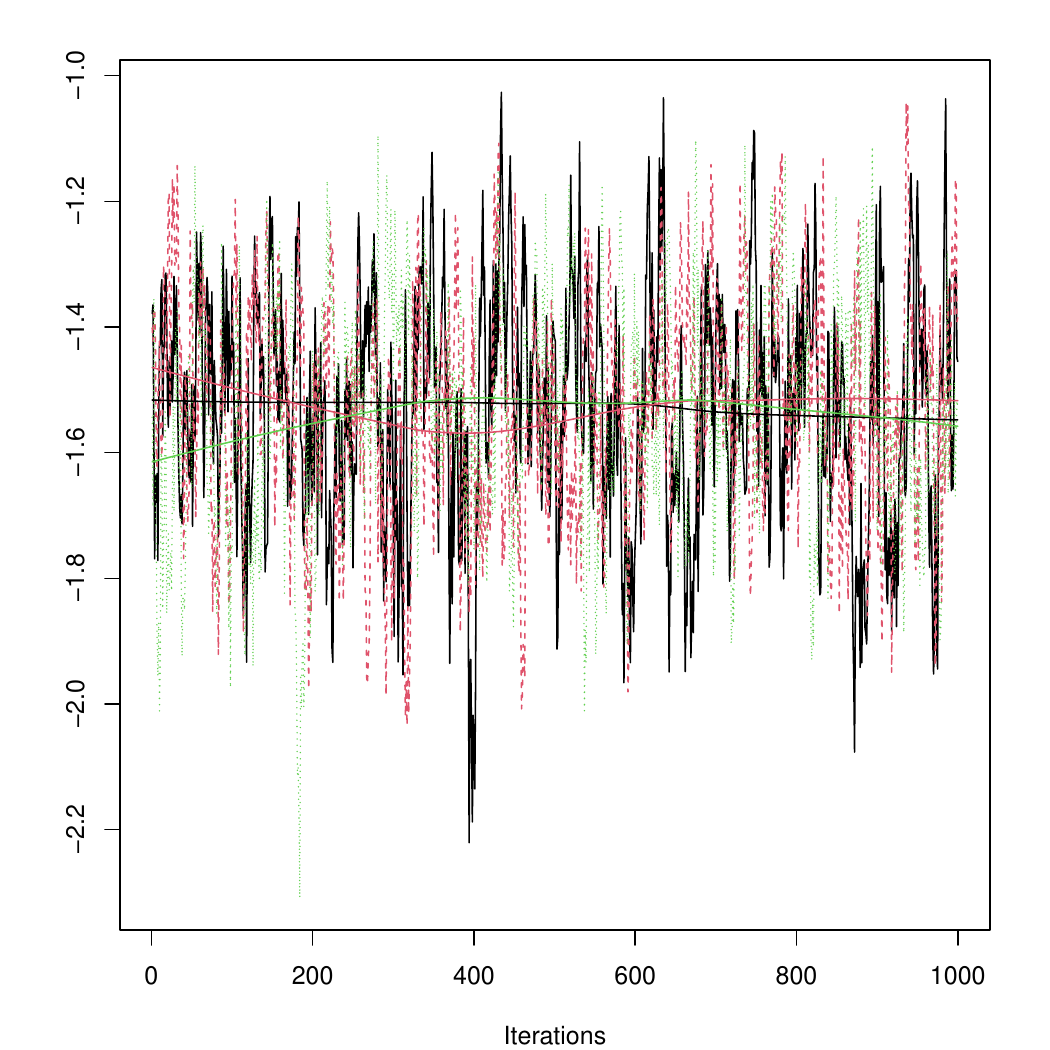}\label{fig:beta1}}
	\subfigure[$\alpha_1$]
{\includegraphics[width=0.32\textwidth]{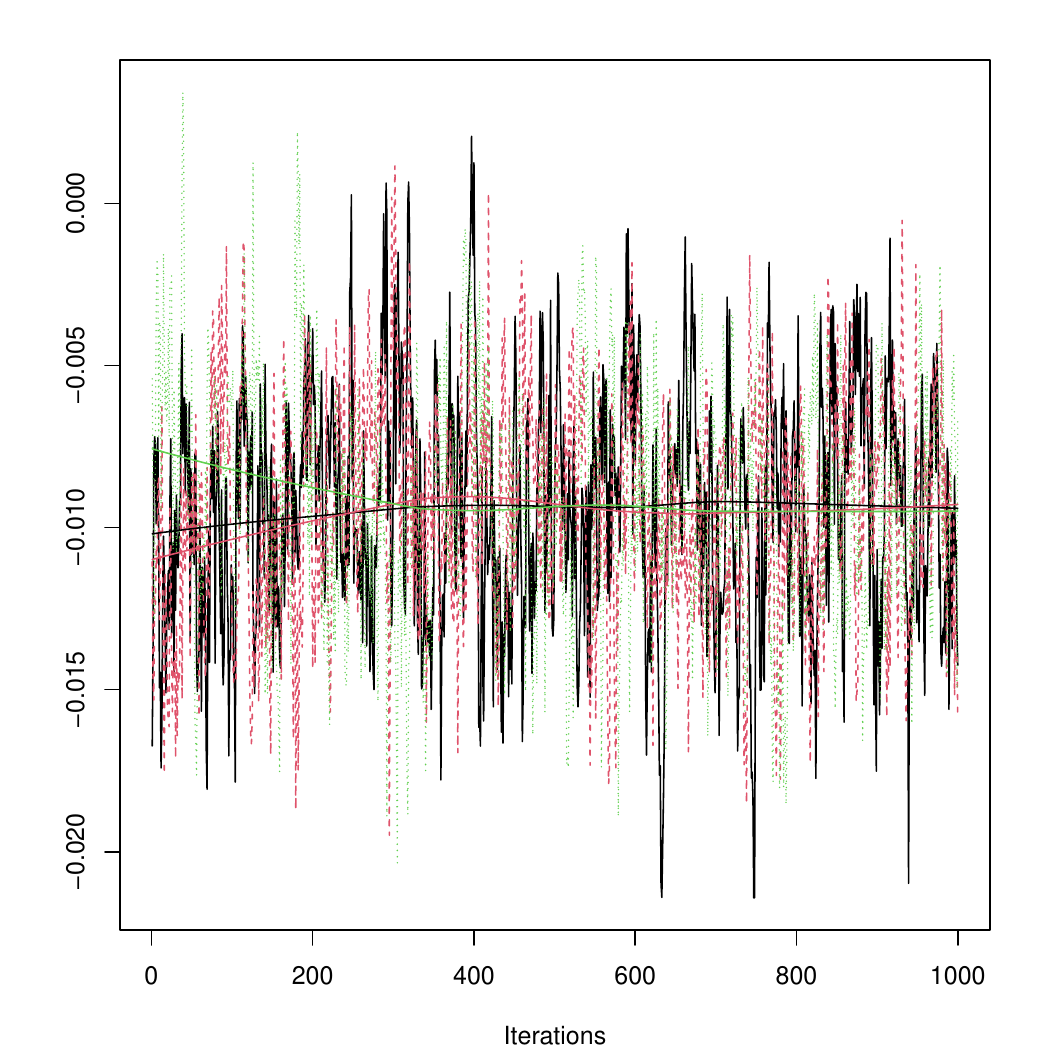}\label{fig:beta2}}
	\subfigure[$\alpha_2$]
{\includegraphics[width=0.32\textwidth]{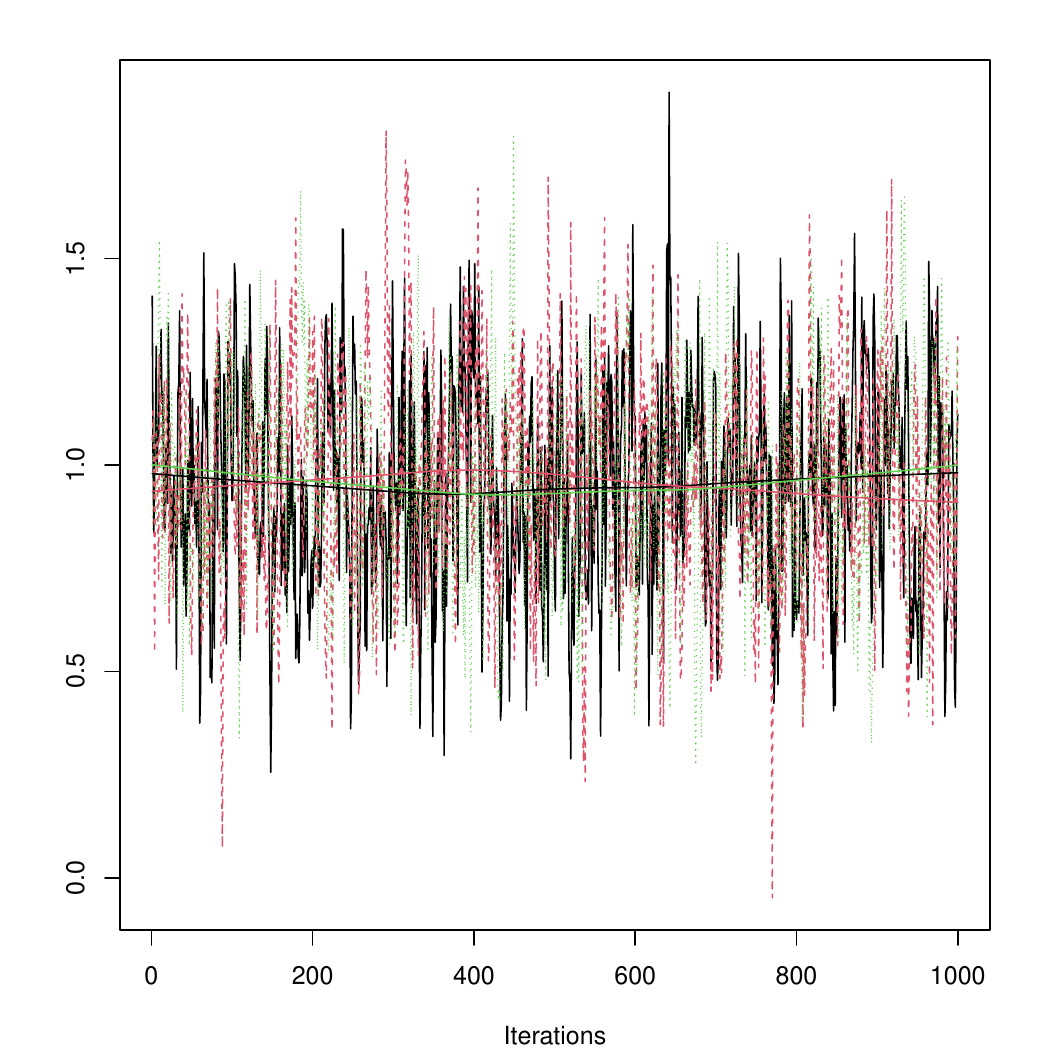}\label{fig:beta3}}
	\caption{Trace plots of parameters in Eq. \eqref{eq:logit}.\label{fig:beta-sim}}
\end{figure}
\begin{figure}[htbp!]
	\centering
	\subfigure[$\beta_0$]
{\includegraphics[width=0.32\textwidth]{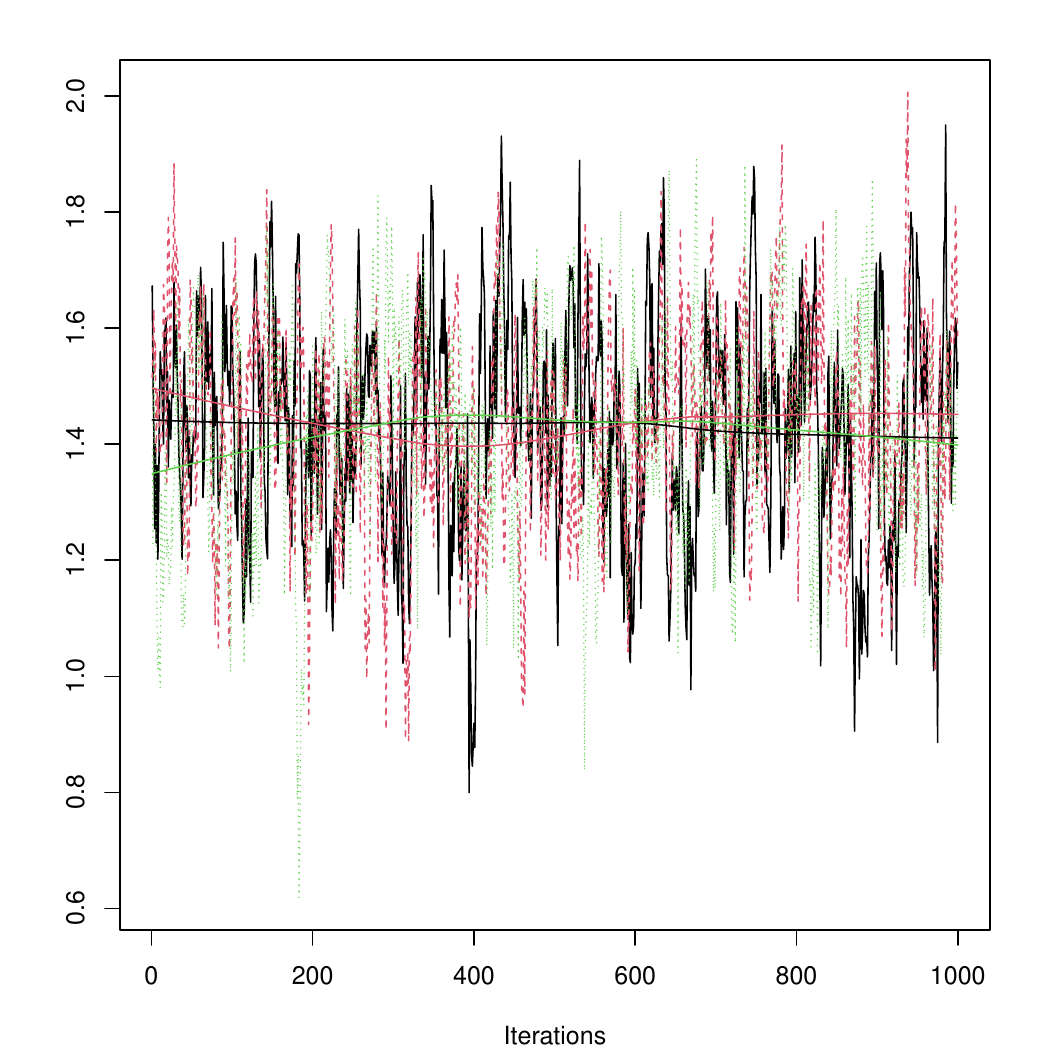}\label{fig:g1}}
	\subfigure[$\beta_1$]
{\includegraphics[width=0.32\textwidth]{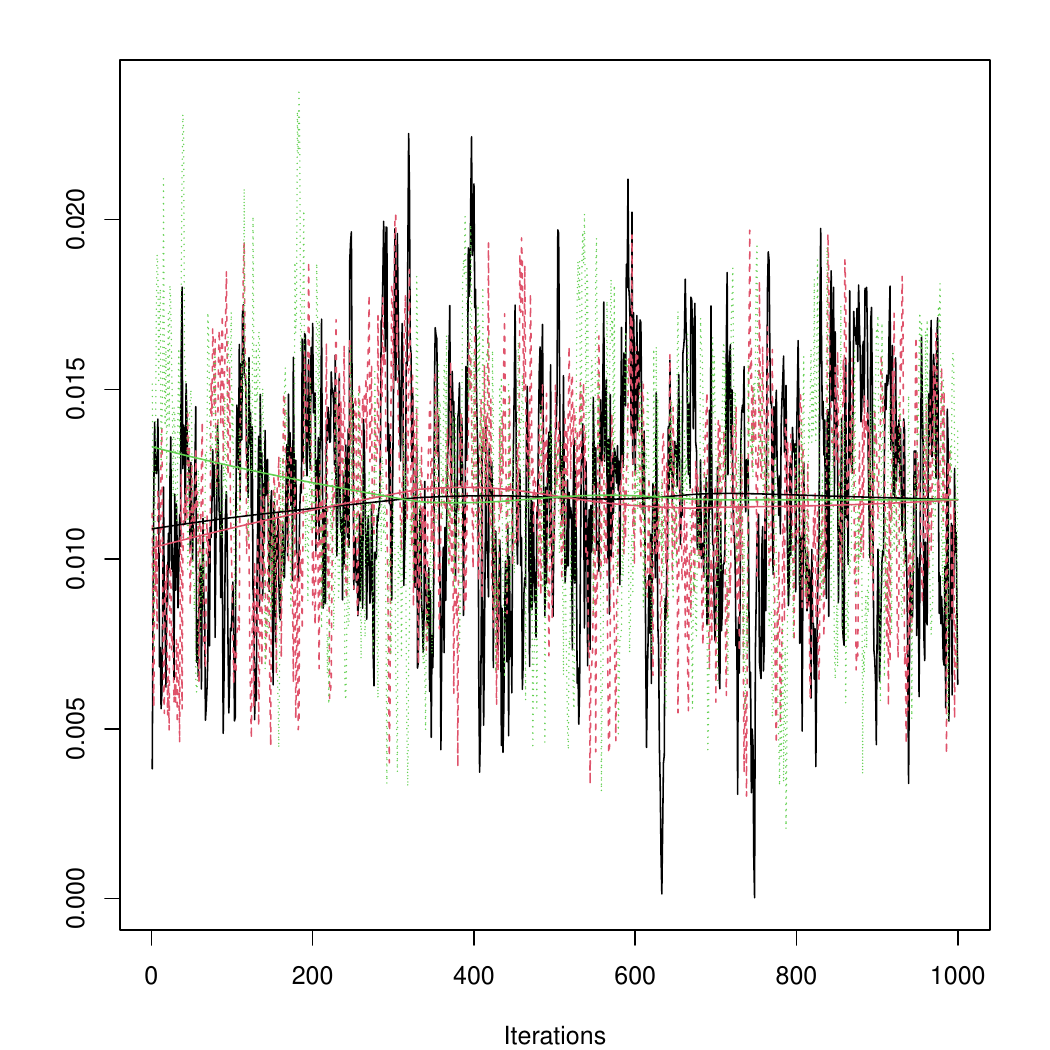}\label{fig:g2}}
	\subfigure[$\beta_2$]
{\includegraphics[width=0.32\textwidth]{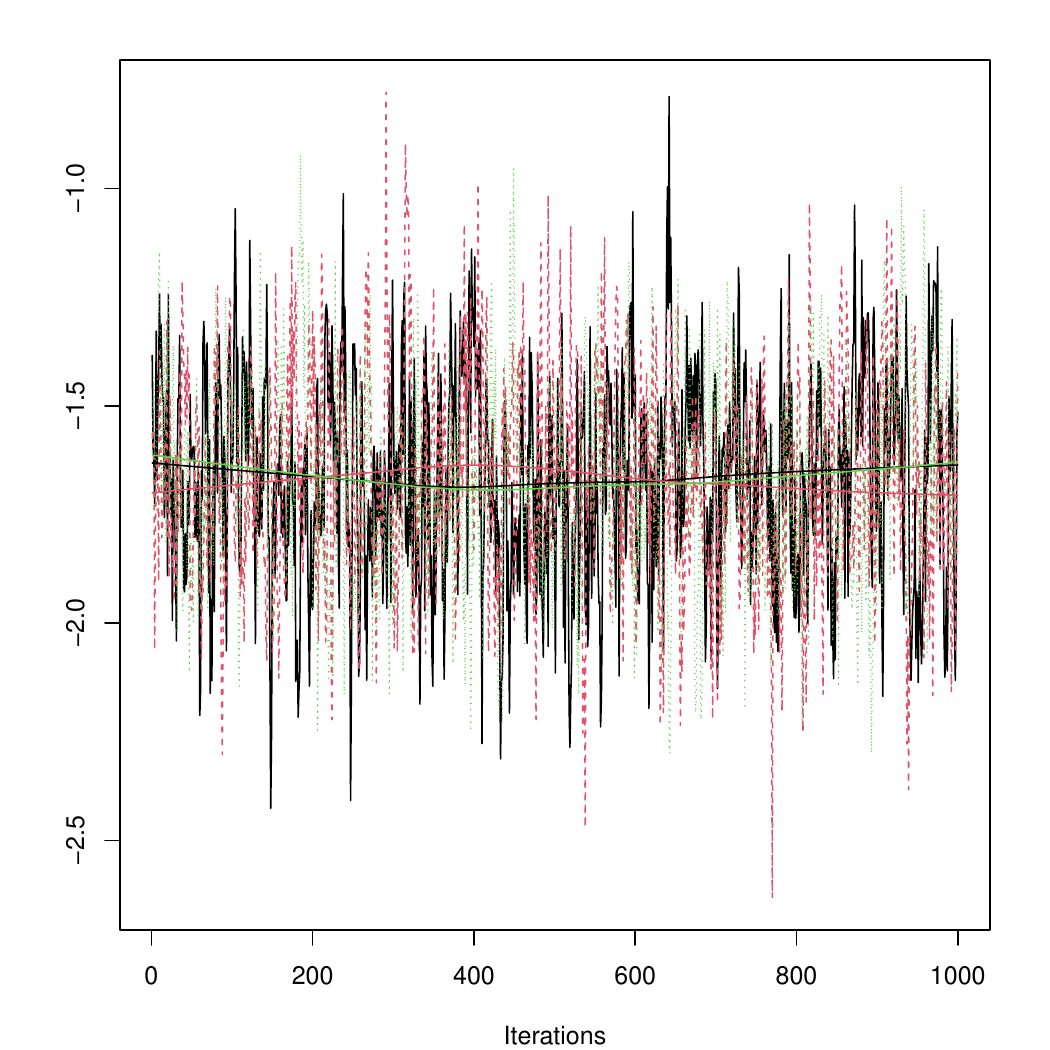}\label{fig:g3}}
	\caption{Trace plots of parameters in Eq. \eqref{eq:rtd}.\label{fig:trace-sim}}
\end{figure}

We use the following setting in the model fitting. 
Three MCMC chains are used to estimate the posterior distributions to ensure 
convergence and assess the robustness of the results. The initial values are randomly generated based on prior distributions. The burn-in sample is set to be $1e+06$, while the number of iterations is  $1.1e+06$. To reduce autocorrelation between samples, we use 100 as the thinning parameter, i.e., every 100th sample will be retained from each chain.

To evaluate the convergence of the Markov chains under consideration, we examine the trace plots for the parameters of Eq. \eqref{eq:logit}  presented in Figure \ref{fig:beta-sim}  and for the parameters of Eq. \eqref{eq:rtd}  shown in Figure \ref{fig:trace-sim}. These plots reveal that all three chains demonstrate patterns indicative of convergence. In addition to visual inspection of the trace plots, we quantitatively assess convergence using the Potential Scale Reduction Factor (PSRF). The PSRF provides a comparative measure of the variance within each chain to the variance across chains, as proposed by Brooks and Gelman \cite{gelman1992inference,brooks1998general}. For all the parameters under study, the PSRF values are uniformly below 1.01, with the multivariate PSRF also recorded at 1.01. These findings provide robust evidence of convergence across the MCMC  chains in our analysis, underscoring the reliability of the sampling process in capturing the underlying statistical properties of the model.
\begin{table}[htb!]
\centering
\caption{Posterior means and 95\% Bayesian credibility intervals for parameters.\label{tab:summary-stats-sim}}
\begin{tabular}{l|cc|c}
\hline
Parameters & Mean & 95\% Credibility Interval &True\\
\hline
$\alpha_0$ & -1.5344 & (-1.9026, -1.2040) & -1.5\\
$\alpha_1$ & -0.0094 & (-0.0164, -0.0026) &-0.01\\
$\alpha_2$ & 0.9545 & (0.4804, 1.4330)&0.8\\
$\beta_0$ & 1.4265 & (1.0636, 1.7550) &1.5 \\
$\beta_1$ & 0.0118 & (0.0053, 0.0184) &0.01\\
$\beta_2$ & -1.6673 & (-2.1317, -1.2123)&-1.8 \\ \hline
$\sigma_1$ & 0.7244 & (0.0497, 2.2164) &--\\
$\sigma_2$ & 1.5081 & (0.1085, 4.7381) &--\\
$\sigma_3$ & 0.6274 & (0.0432, 2.0101) &--\\
$\sigma_4$ & 0.7031 & (0.0492, 2.1804) &--\\
$\sigma_5$ & 1.5195 & (0.1151, 4.8460) &--\\
$\sigma_6$ & 1.0358 & (0.0713, 3.2423) &--\\
\hline
\end{tabular}
\end{table}

Table \ref{tab:summary-stats-sim} presents the posterior means and corresponding 95\% Bayesian credibility intervals for the estimated parameters. We observe that the posterior means demonstrate a close proximity to the true values, indicating a robust performance of the estimation procedure. Furthermore, all 95\% Bayesian credibility intervals encompass the true parameter values, signifying the reliability and accuracy of the Bayesian inference process.
\begin{figure}[htb!]
	\centering
	\includegraphics[width=0.5\textwidth]{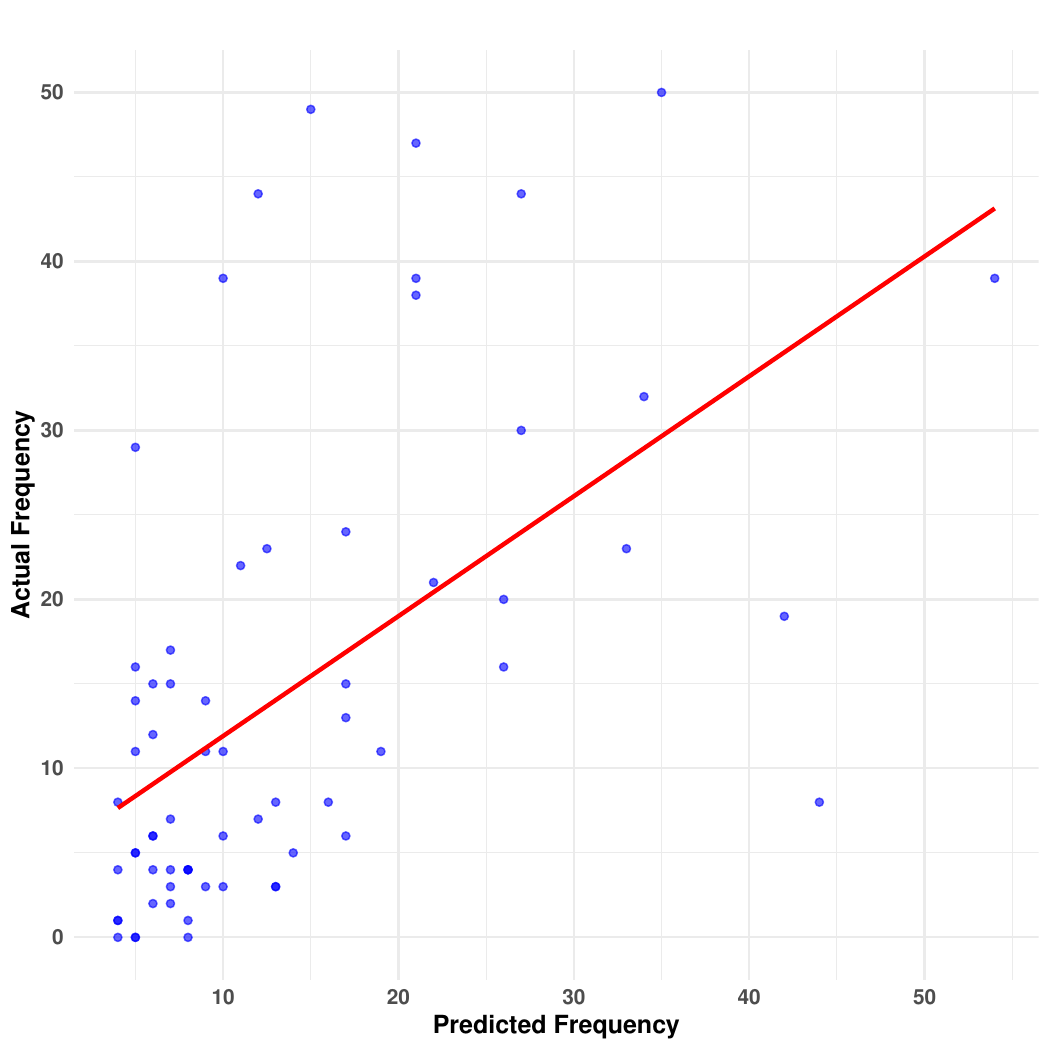}
	\caption{Nowcast the IBNR cyber incidents.  \label{fig:IBNR-pred-syn}}
\end{figure}

An advantage of employing the Bayesian approach is the direct simulation of nowcasts from the model itself. In our analysis, we utilize the simulated median as the prediction. Figure \ref{fig:IBNR-pred-syn} illustrates a scatter plot comparing the nowcasted frequencies with the real frequencies. Upon inspection, it becomes apparent that there are instances where the model underpredicts the frequencies. One possible explanation for this discrepancy could be the presence of outliers in the synthetic data, as observed in Figure \ref{fig:delay-freq-sim}. Despite the presence of outliers, the Pearson correlation coefficient of 0.548 suggests a reasonably strong positive correlation between the nowcasted and real frequencies. This correlation indicates that the proposed model demonstrates good predictive performance when applied to synthetic data, even in the presence of outliers.

 \section{Empirical study}\label{sec:empirical}
In this section, we employ the proposed model to the ITRC data. Recall that the in-sample data is from 1/1/2018 to 12/31/2022, while the out-of-sample data is from 1/1/2023 to 12/31/2023.
 
\subsection{Model fitting}
In our Bayesian model, we employ three chains to estimate posterior distributions, initializing them with random values drawn from standard exponential distributions based on prior information. We set the burn-in sample to $2 \times 10^6$, with a total number of iterations at $2.1 \times 10^6$. Again, we utilize a thinning parameter of 100 to ensure efficient sampling.

\begin{figure}[htbp!]
	\centering
	\subfigure[$\alpha_0$]
{\includegraphics[width=0.32\textwidth]{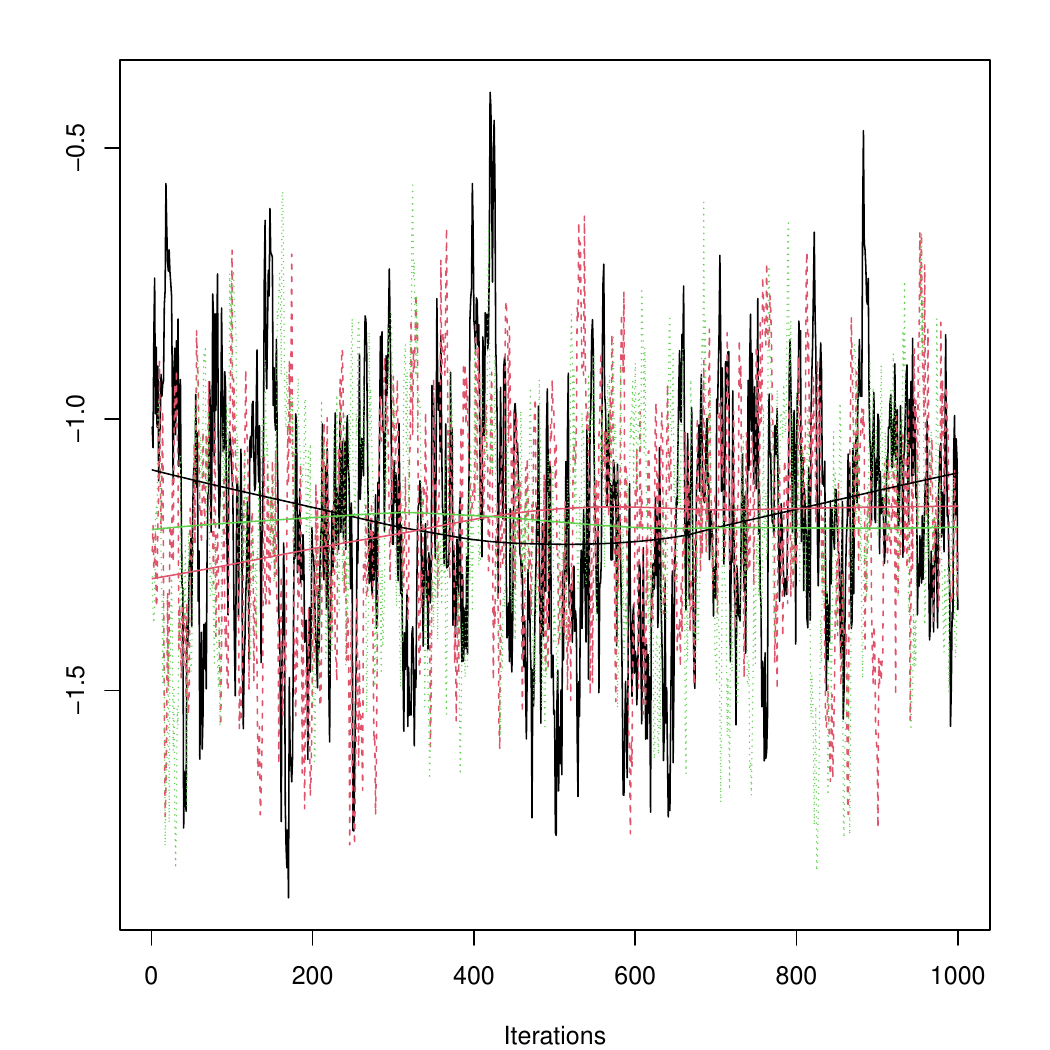}\label{fig:beta1}}
	\subfigure[$\alpha_1$]
{\includegraphics[width=0.32\textwidth]{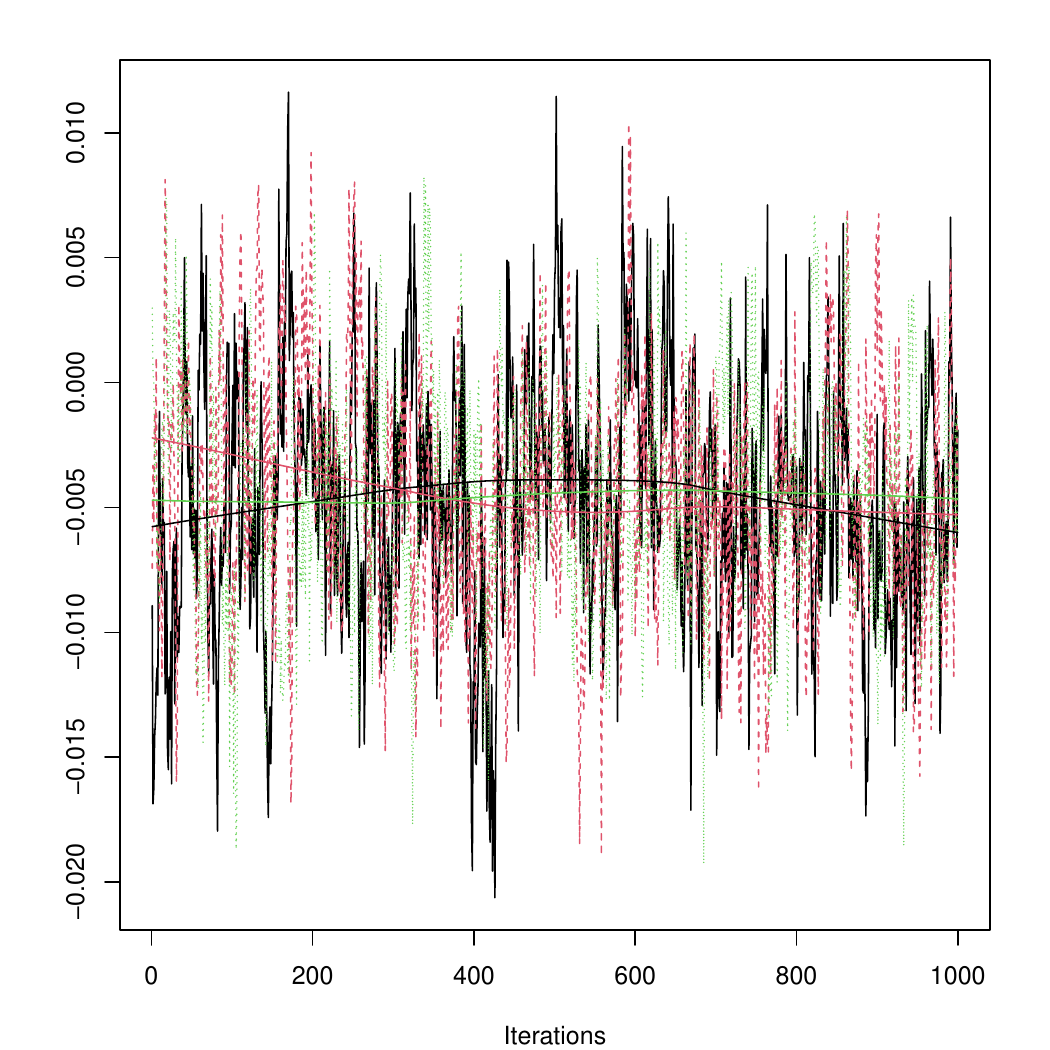}\label{fig:beta2}}
	\subfigure[$\alpha_2$]
{\includegraphics[width=0.32\textwidth]{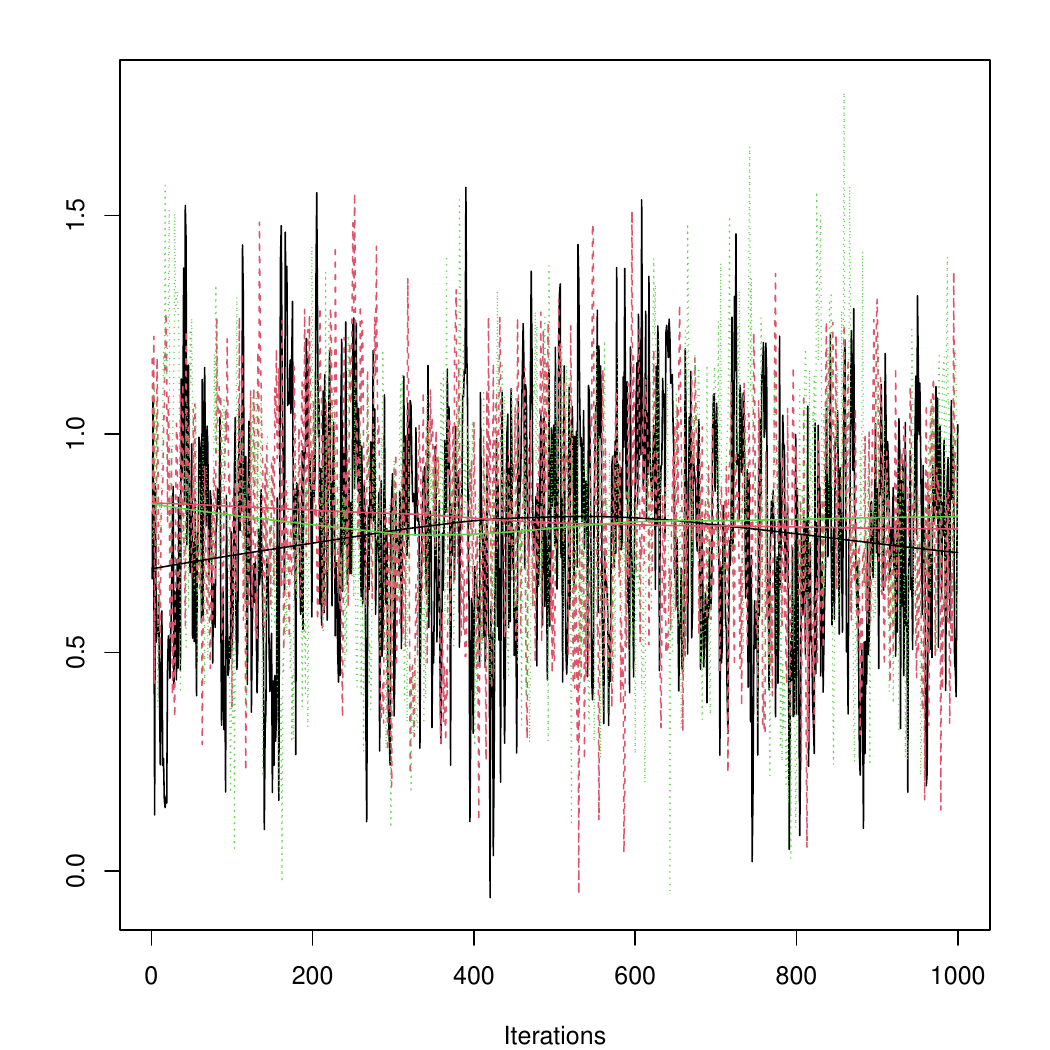}\label{fig:beta3}}
	\caption{Trace plots of parameters in Eq. \eqref{eq:logit}.\label{fig:beta}}
\end{figure}

\begin{figure}[htbp!]
	\centering
	\subfigure[$\beta_0$]
{\includegraphics[width=0.32\textwidth]{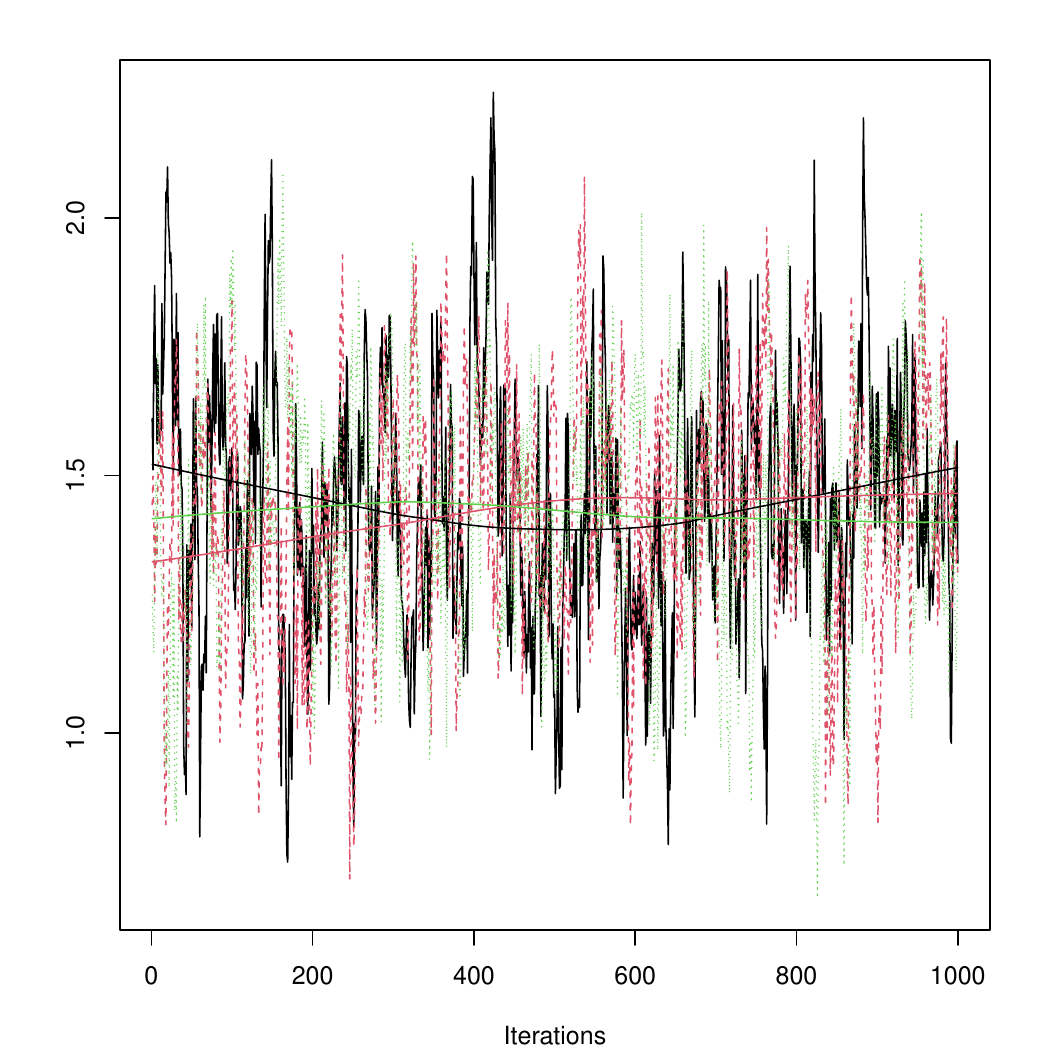}\label{fig:g1}}
	\subfigure[$\beta_1$]
{\includegraphics[width=0.32\textwidth]{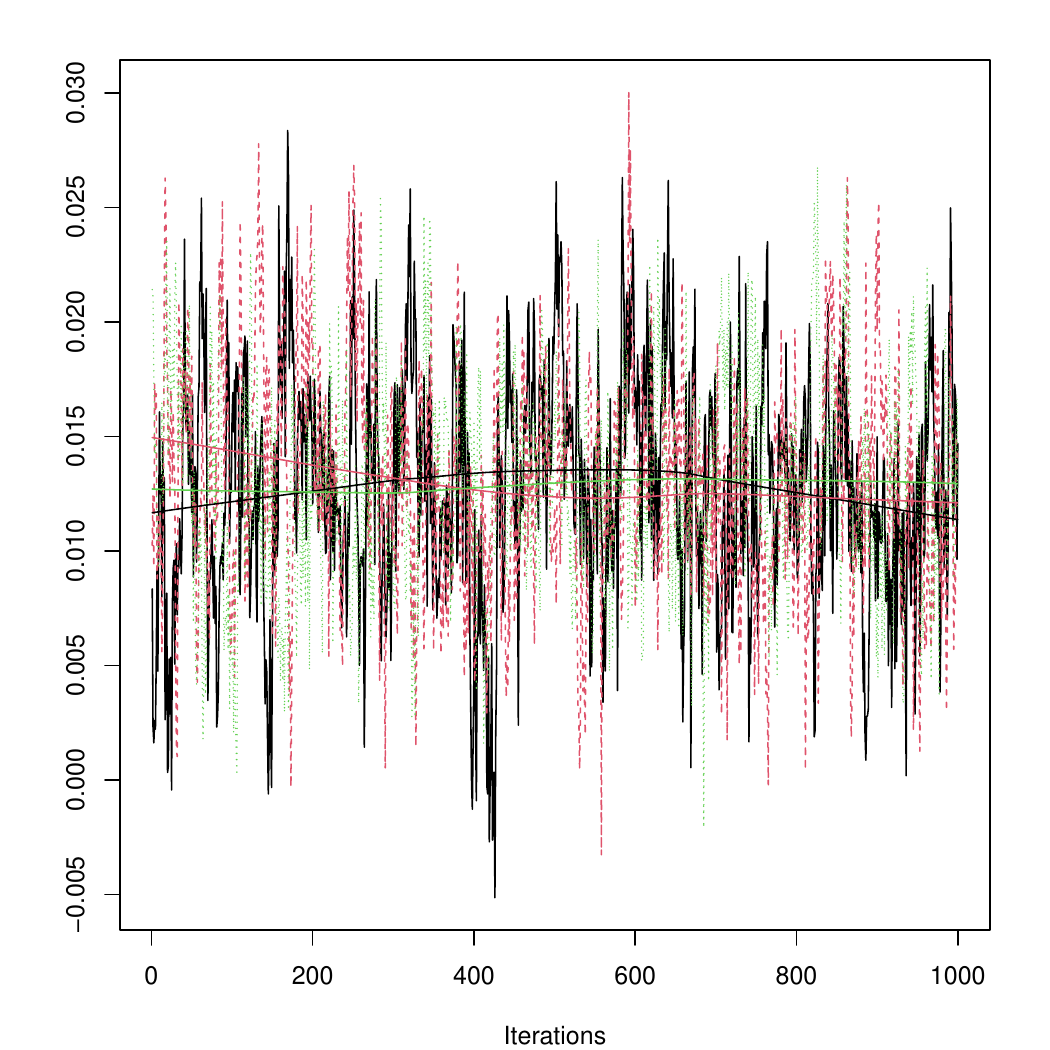}\label{fig:g2}}
	\subfigure[$\beta_2$]
{\includegraphics[width=0.32\textwidth]{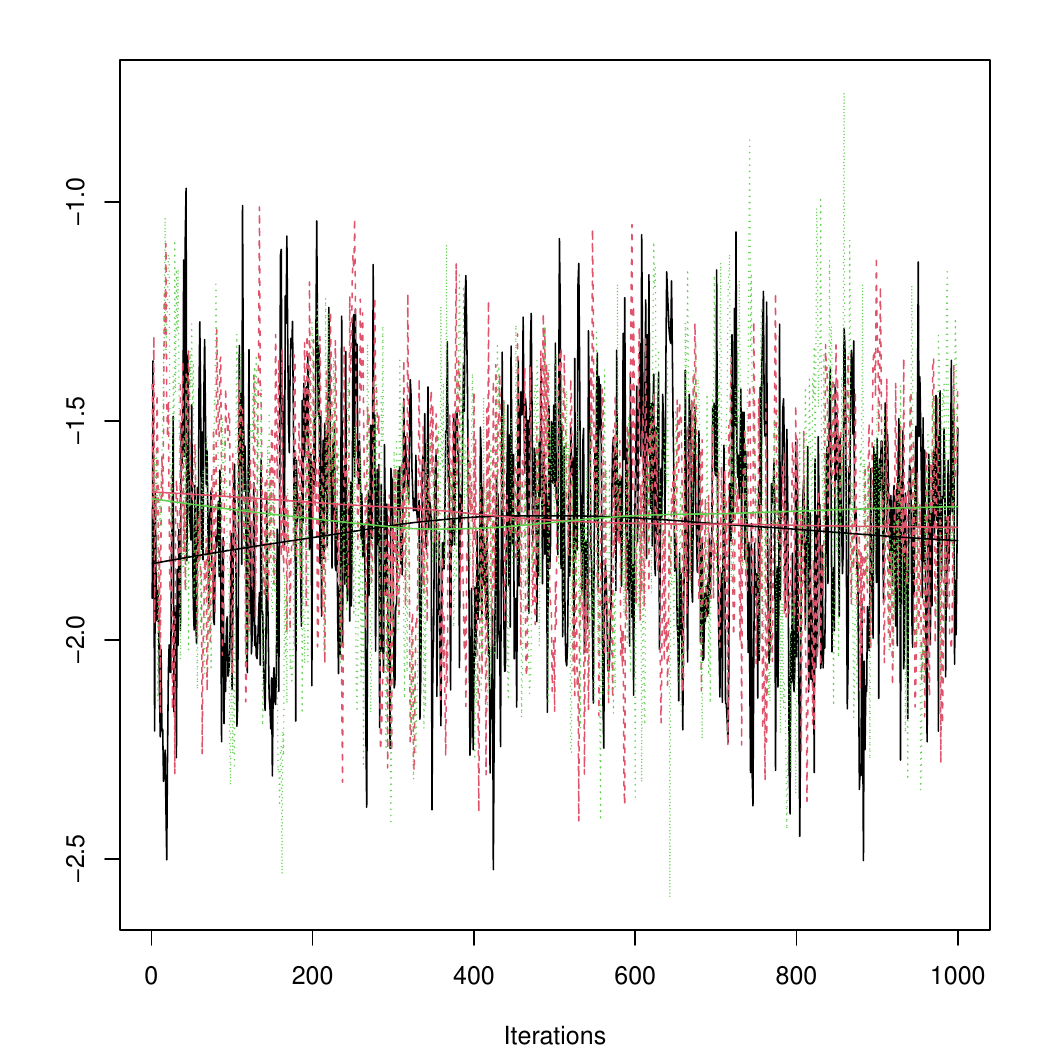}\label{fig:g3}}
	\caption{Trace plots of parameters in Eq. \eqref{eq:rtd}.\label{fig:trace}}
\end{figure}

Examining the trace plots for the parameters depicted in Figure \ref{fig:beta} and Figure \ref{fig:trace}, it is evident that all three chains exhibit patterns suggestive of convergence. These trace plots provide a visual representation of how the parameter values evolve over the course of the sampling process. The convergence patterns observed in the trace plots indicate that the chains have reached stable distributions, suggesting that the MCMC algorithm has successfully explored the posterior space and obtained reliable estimates of the model parameters. 
 
For all parameters investigated, the PSRF values consistently register below 1.01, with the multivariate PSRF also measured at 1.01. These results further offer compelling evidence of convergence across the MCMC chains utilized in our analysis.

\begin{table}[htb!]
\centering
\caption{ Posterior means and 95\% Bayesian credibility intervals for parameters.}
\label{tab:parameters}
\begin{tabular}{l|cc}
\hline
Parameter & Mean & 95\% Credibility Interval \\ 
\hline
$\alpha_0$ & -1.1895 & (-1.6363, -0.7578) \\
$\alpha_1$ & -0.0045 & (-0.0139, 0.0050) \\
$\alpha_2$ & 0.7907 & (0.2664, 1.2950) \\
$\beta_0$ & 1.4315 & (0.9691, 1.8858) \\
$\beta_1$ & 0.0129 & (0.0031, 0.0226) \\
$\beta_2$ & -1.7282 & (-2.2510, -1.2183) \\ \hline
$\sigma_{\alpha_0}$ & 0.7220 & (0.0515, 2.3004) \\
$\sigma_{\alpha_2}$ & 1.5221 & (0.1031, 4.9042) \\
$\sigma_{\alpha_3}$ & 0.6032 & (0.0472, 1.9400) \\
$\sigma_{\beta_0}$ & 0.8610 & (0.0555, 2.7244) \\
$\sigma_{\beta_1}$ & 1.5246 & (0.1277, 4.6604) \\
$\sigma_{\beta_2}$ & 1.1488 & (0.0743, 3.7174) \\
\hline
\end{tabular}
\end{table}
Table \ref{tab:parameters} displays the posterior mean and 95\% Bayesian credibility intervals for the parameters within our proposed model. The parameter estimates $\alpha_1$  and $\alpha_2$  exhibit varying degrees of influence on the model's dynamics. Specifically, $\alpha_1$ presents a negative impact, suggesting an increased difficulty in system breaches over recent years. In contrast, the positive effect of $\alpha_2$ indicates that, when breaches do occur, their discovery might be delayed, reflecting potentially slower detection processes or more sophisticated breach tactics. Further analysis of the $\beta$ parameters reveals additional insights. $\beta_1$  shows a positive effect, suggesting an increase in the frequency of incidents in recent years. Conversely, $\beta_2$  indicates a negative effect, which could be interpreted as a reduction in the number of incidents with large delays before discovery. While the mean estimate for $\alpha_1$  might initially suggest a minimal impact on the model, its inclusion enhances the model's predictive accuracy. This highlights the nuanced role $\alpha_1$ plays in the model, potentially capturing dynamics in system breach incidents and their detection timelines. Therefore, despite its seemingly minor direct impact, we include $\alpha_1$  in our analysis.  
 
\subsection{Nowcasting performance}
 In our model, the median of simulated samples is used as the prediction. 
\begin{figure}[htbp!]
	\centering
	\includegraphics[width=0.6\textwidth]{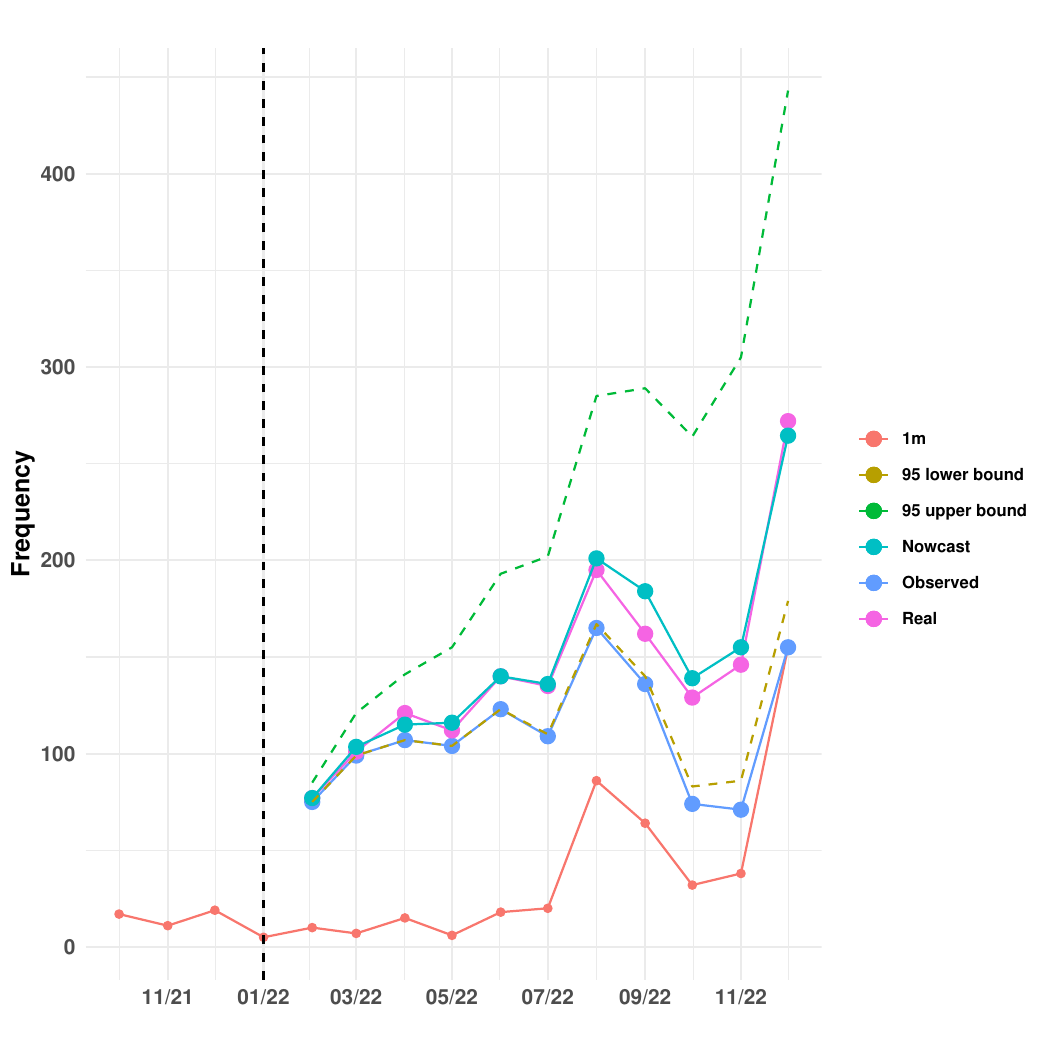}
	\caption{Nowcast the IBNR cyber incidents. The  `1m' line represents the number of cyber incidents that are reported during the occurring month. The dashed lines of `95 lower bound' and `95 upper bound' represent the 95\% prediction bounds based on the proposed model. The observed number of incidents 
represents all the incidents that have been observed until December 2022. \label{fig:IBNR-pred}}
\end{figure}
Figure \ref{fig:IBNR-pred} presents the nowcasting outcomes accompanied by 95\% prediction intervals, juxtaposed with the observed incident counts. The `1m' line reveals a scant number of incidents reported during the respective occurring months. The observed incidents for any given month reflect the cumulative count of occurrences discovered until the end of December 2022. For example, in January 2022, 5 incidents were reported, while February 2022 saw 18 reported incidents, all of which had transpired in January 2022. Thus, the total number of incidents occurring in January 2022 and reported by the end of December 2022 amounted to 91. Similarly, February 2022 saw 10 reported incidents, yet the total observed incidents for that month by the end of December 2022 stood at 75. The nowcast line delineates the total incidents occurring in a specific month and reported within 12 months. For instance, the total number of incidents occurring in February 2022 and reported within 12 months, as estimated by the proposed model, was 77. Conversely, the `real' line portrays the total incidents occurring in a specific month and observed within 12 months, based on testing data. In the case of February 2022, the total observed incidents by the end of January 2023 amounted to 77. Figure \ref{fig:IBNR-pred} illustrates a close alignment between the nowcast and real incident counts during the initial months, with slight deviations in subsequent months. This discrepancy can be attributed to the reduced number of observations in later months, which is evident from the wider prediction intervals.

%In summary, the proposed model demonstrates satisfactory nowcasting performance, effectively estimating incident counts with notable accuracy, particularly in the earlier periods.

\begin{figure}[htb!]
	\centering
	\includegraphics[width=0.5\textwidth]{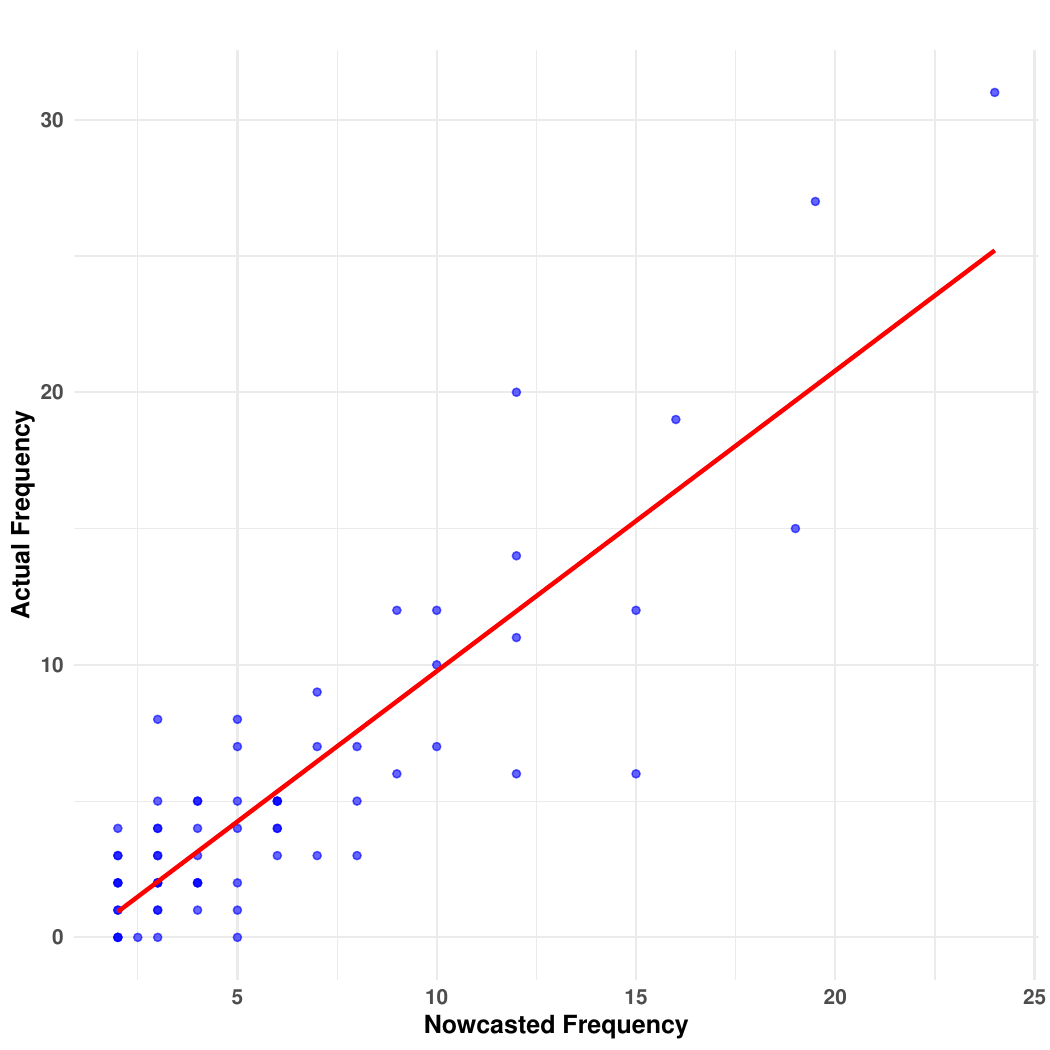}
	\caption{Nowcast the IBNR cyber incidents.  \label{fig:IBNR-pred-real}}
\end{figure}
Similarly, we apply the Bayesian approach to nowcast the delayed incidents, utilizing the simulated median as the prediction. Figure \ref{fig:IBNR-pred-real} presents a scatter plot comparing the nowcasted frequencies with the real frequencies. Remarkably, a high correlation between the two is evident, with the Pearson correlation coefficient reaching 0.875. This substantial correlation suggests that the proposed model demonstrates robust predictive performance.
%The strong positive correlation observed between the nowcasted and real frequencies underscores the accuracy and effectiveness of the Bayesian approach in forecasting delayed incidents.  

\paragraph{Model comparison.} To further assess the model performance, we compare the proposed model (labeled as M0) with other commonly used ones in the literature. 

\begin{itemize}
	\item M1: Chain-ladder model \cite{barnett2000best}. This is the most commonly used approach in practice.   It assumes that the proportional developments of claims from one development period to the next are the same for all origin periods.
\item M2: Mack chain-ladder model  \cite{mack1993distribution,mack1999the}. 
It is a distribution-free stochastic model underlying the  Chain-ladder approach. Let $S_{ik}$ denote the cumulative claims of origin period (e.g. accident year) $i=1,\ldots,m$, with claims known for development period (e.g. development year) $1\leq k\leq n+1-i$.
	In order to forecast the amounts $S_{ik}$ for $n+1-i< k\leq n$, the Mack chain-ladder model assumes:
	\begin{equation*}
		\begin{aligned}
			&{\rm E}[F_{i,k}|S_{i1},S_{i2},\ldots S_{ik}]=f_k~{\rm with}~F_{i,k}=\frac{S_{i,k+1}}{S_{ik}},\\ 
			&{\rm Var}[F_{i,k}|S_{i1},S_{i2},\ldots S_{ik}]=\frac{\sigma_{k}^2}{w_{ik}S_{ik}^\alpha},\\ 
			&(S_{i1},S_{i2},\ldots S_{in}) ~\mbox{are independent by}~i,~1\leq i\leq n
		\end{aligned}
	\end{equation*}
	with $w_{ik}\leq [0,1]$, $\alpha\in \{0,1,2\}$.  
	
	\item M3: Bootstrap chain-ladder model \cite{england1999analytic}.  It employs a two-stage bootstrapping methodology to forecast the trajectory of the development run-off. In the initial stage, the model applies the ordinary chain-ladder method to the cumulative claims triangle. Subsequently, scaled Pearson residuals are computed from this stage, and through bootstrapping iterations ($R=1000$ times in our study), future incremental claims payments are projected using the standard chain-ladder method. Moving to the second stage, the process error is simulated by incorporating the bootstrapped values as means, while adhering to the assumed process distribution. The resulting set of reserves derived from this approach constitutes the predictive distribution. For the bootstrap chain-ladder model, we use the median as the prediction.
	
		\item M4: Generalized linear model. Renshaw and Verrall \cite{renshaw1998stochastic} proposed a methodology for modeling incremental claims utilizing an "over-dispersed" Poisson distribution. If we denote the incremental claims for origin year $i$ in development year $k$ as $C_{ik}$, the model is characterized by the following equations:
\begin{equation*}\label{eqn:glm_pois}
\mathbb{E}[C_{ik}] = m_{ik}, \quad \text{and} \quad \text{Var}[C_{ik}] = \phi \mathbb{E}[C_{ik}] = \phi m_{ik},
\end{equation*}
where  $\phi$ is a scale parameter, $\log(m_{ik}) = \eta_{ik}$, and $\eta_{ik}= c + \alpha_i + \beta_k$, with $\alpha_1= \beta_1 = 0$. It is a generalized linear model wherein the response is modeled using a logarithmic link function, and the variance is directly proportional to the mean.

\end{itemize}

Model comparison based on root Mean Squared Error (MSE) and Mean Absolute Error (MAE) is a common approach to assessing the predictive performance of different models. MSE measures the average squared difference between predicted values and actual values, while MAE measures the average absolute difference between predicted values and actual values.

\begin{table}[htb!]
\begin{center}
    \caption{Prediction results of IBNR incidents from various models.\label{tab:pred}}
\begin{tabular}{l|ccccc}
	\hline
	& M0 & M1 & M2 & M3 & M4 \\
	\hline
	 RMSE	&  {\bf 3.0063}  & 26.0646 &7.2199&7.4944 &16.6219\\
	 MAE	&  {\bf 2.2727} & 14.4414 & 5.7634   & 5.4091   & 8.4394 \\
	\hline
\end{tabular}
\end{center}
\end{table}
Referring to Table \ref{tab:pred}, which displays the prediction results from various models, it is evident that the performance of the models varies significantly. The proposed model M0 exhibits the lowest RMSE and MAE, indicating superior predictive accuracy compared to the other models. Conversely, Model M1 demonstrates the highest RMSE and MAE values, suggesting poorer predictive performance. Models M2 and M3 fall within intermediate ranges of RMSE and MAE values, while Model M4  has a poor prediction performance. 

In summary, the proposed model demonstrates satisfactory nowcasting performance, effectively estimating incident counts with high accuracy. 

\subsection{Reserve Estimation}
We apply the results derived from our proposed Bayesian model to estimate the financial impact of cyber incidents, leveraging the average cost of a data breach. According to IBM's 2022 report, the average cost of a data breach incident is estimated at \$4.35 million. This value serves as the per-incident cost in our analysis. By integrating this cost metric with the model's now-cast incident frequency, we can approximate the total financial impact and the insurance reserves that are expected to cover over the year.

Table \ref{tab:reserve-estimation} presents the reserve estimation for cyber incidents in 2022, employing the proposed Bayesian model. We calculate monthly figures for estimated, paid, IBNR, and ultimate claims, all denominated in millions based on the nowcast frequencies, where the ultimate claims are computed by using the out-of-sample data of  2023. A consistent trend observed over the year is the widening discrepancy between estimated and paid claims, leading to a significant accumulation of IBNR claims. The peak in estimated claims was witnessed in December, surging to 1150.575 million, alongside a notable increase in IBNR claims to 476.325 million. It is interesting to see that the IBNR rate shows an overall decreasing trend. The comparative analysis of estimated versus ultimate claims underscores the good performance of the proposed Bayesian model in managing the uncertainties in cyber incident claims. With estimated total claims reaching 7490.7 million, slightly above the ultimate claims of 7312.35 million, the model exhibits a marginally conservative bias. Nevertheless, this nuance further underscores the model's significant promise in refining actuarial methodologies for cyber risk management.

\begin{table}[htbp!]
\centering
\caption{Reserve estimation for cyber incidents in 2022 with IBNR Change Rate. The \textbf{IBNR(\%)}  represents the IBNR change rate. Unit: million.}
\label{tab:reserve-estimation}
\begin{tabular}{@{}lccccc@{}}
\toprule
\textbf{Month} & \textbf{Estimated Claims} & \textbf{Paid Claims} & \textbf{IBNR Claims} & \textbf{Ultimate Claims} & \textbf{IBNR(\%)} \\ \midrule
01/22 & 395.850 & 395.85 & 0.000 & 395.85 & -- \\
02/22 & 334.950 & 326.25 & 8.700 & 334.95 & --\\
03/22 & 450.225 & 430.65 & 19.575 & 439.35 & 125.00 \\
04/22 & 500.250 & 465.45 & 34.800 & 526.35 & 77.78 \\
05/22 & 504.600 & 452.40 & 52.200 & 487.20 & 50.00 \\
06/22 & 609.000 & 535.05 & 73.950 & 609.00 & 41.67 \\
07/22 & 591.600 & 474.15 & 117.450 & 587.25 & 58.82 \\
08/22 & 874.350 & 717.75 & 156.600 & 848.25 & 33.33 \\
09/22 & 800.400 & 591.60 & 208.800 & 704.70 & 33.33 \\
10/22 & 604.650 & 321.90 & 282.750 & 561.15 & 35.42 \\
11/22 & 674.250 & 308.85 & 365.400 & 635.10 & 29.23 \\
12/22 & 1150.575 & 674.25 & 476.325 & 1183.20 & 30.36 \\  \midrule
\textbf{Total} & \textbf{7490.7} & \textbf{5694.15} & \textbf{1796.55} & \textbf{ 7312.35}& \\ \bottomrule
\end{tabular}
\end{table}

\section{Conclusion}\label{sec:conclusion}

This work introduces a novel Bayesian nowcasting model tailored to estimate IBNR cyber incidents. Through the synthetic and empirical data studies, the model has demonstrated exceptional predictive accuracy, outperforming traditional models commonly used in actuarial science, such as the chain-ladder and generalized linear models. The application of our model in reserve estimation, utilizing the average cost of a data breach reported by IBM, offers a refined approach to quantifying the financial implications of cyber incidents. This not only provides reserves estimations but also contributes valuable insights into the dynamic nature of cyber risk exposure. The proposed Bayesian nowcasting model represents a significant advancement in actuarial science, particularly in the domain of cyber risk management. Its ability to effectively integrate temporal and correlation effects among delayed reporting intervals sets a new benchmark for predictive analytics in insurance reserves.

While this study focuses on cyber incidents, the proposed model is adaptable to other types of insurance claims where reporting delays are a concern. By integrating more detailed data regarding the specifics of cyber incidents, the model's predictive capabilities can be enhanced. For instance, in the context of reserve estimation, employing actual cost data for each incident, rather than relying on average costs, would likely yield more precise outcomes. Nonetheless, obtaining such detailed cost data is often challenging in practice, presenting a notable limitation to this approach. Furthermore, exploring a joint modeling approach that simultaneously considers both the frequency of incidents and their associated costs promises to refine the model's performance even further.

% Given that there exists no universal methodology to convert the quantity of breached records to a dollar loss, our strategy is to designate the number of breached records as the loss directly. In underwriting scenarios, dollar loss might be determined by referring to the historical mean loss per breached record. For example, IBM \cite{IBM2022} reported the average cost of a data breach per record to be \$164 in 2022. 

% We convert the severity to the dollar loss using a formula adopted from the literature \cite{romanosky2016examining} and assess the
% scoring performance based on the dollar loss (DL).
% $$\log({\rm DL})=7.68+0.76*\log(LBR).$$
 
 %\input{delay-submitted.bbl}
\bibliographystyle{plain}  % more readable regarding references
\bibliography{breach.bib,breach_sun.bib}
\end{document}